\documentclass[aps,prx,reprint,superscriptaddress,showpacs,showkeys, nofootinbib]{revtex4-2}

\usepackage{amsmath,amssymb,bm}
\usepackage{graphicx}
\usepackage[caption=false]{subfig} % 'subcaption' can conflict with revtex; using 'subfig'
\usepackage{hyperref}
\usepackage{amssymb}

\usepackage[dvipsnames]{xcolor}
\definecolor{forestgreen}{rgb}{0.13, 0.55, 0.13}
\usepackage[normalem]{ulem}

\usepackage{soul}
\usepackage{comment}
\usepackage[english]{babel}

\newcommand{\Tss}{t_L} % time for dw to reach end of segment
\newcommand{\Tfp}{T_0} % time of first pause
\newcommand{\xfp}{x_f} % first pausing position
\newcommand{\CNi}{N_\text{i}} % initial cluster length after first pause
\newcommand{\Cmax}{N_{max}} % maximal length of intitial clutser
\begin{document}

\title{Paused in translation: A model for the transcript length-dependent impact of ribosome-targeting antibiotics}

\author{Johannes Keisers}
\email{johannes.keisers@umontpellier.fr}
\affiliation{Centre de Biologie Structurale (CBS), Univ Montpellier, CNRS, INSERM, Montpellier 34090, France}

\author{Norbert Kern}
\affiliation{Laboratoire Charles Coulomb (L2C), Université de Montpellier, CNRS, Montpellier, France}

\author{Luca Ciandrini}
\email{luca.ciandrini@umontpellier.fr}
\affiliation{Centre de Biologie Structurale (CBS), Univ Montpellier, CNRS, INSERM, Montpellier 34090, France}
\affiliation{Institut Universitaire de France (IUF)}

\date{July 25, 2025}

\begin{abstract}
Ribosome-targeting antibiotics, such as chloramphenicol, stall elongating ribosomes during protein synthesis, disrupting mRNA translation. These antibiotic-induced pauses occur stochastically, alter collective ribosome dynamics and transiently block protein production on the affected transcript. 
Existing models of ribosome traffic often rely on idealized assumptions, such as infinitely long mRNAs and simplified pausing dynamics, overlooking key biological constraints. 
Here, we develop a Totally Asymmetric Simple Exclusion Process (TASEP) that incorporates stochastic particle pausing, using experimentally determined pausing and unpausing rates to model the effects of ribosome-targeting antibiotics.
We introduce a Single-Cluster approximation, which is analytically treatable, tailored to capture the biologically relevant regime of rare and long antibiotic-induced pauses.
This biologically constrained model reveals three key insights: (i) the inhibition of antibiotic-induced translation strongly depends on transcript length, with longer transcripts being disproportionately affected; (ii) reducing ribosome initiation rates significantly mitigates antibiotic vulnerability; and (iii) inhibition of translation is governed more by collective ribosome dynamics than by single-ribosome properties.
Our analytical predictions match Gillespie simulations, align quantitatively 
with experimental observations, and yield testable hypotheses for future experiments.
These findings may have broader implications for the mechanistic modeling 
of other biological transport processes (e.g., RNAP dynamics), 
and more generally for the community studying traffic models.

\end{abstract}

\maketitle

\section{Introduction}

The conversion of genetic information into functional proteins is a two-step process requiring precise coordination: the transcription of DNA into messenger RNA (mRNA) by RNA polymerases (RNAPs), followed by the translation of mRNA into proteins by ribosomes. However, because neither RNAPs nor ribosomes can overtake while moving along their respective templates, they are inherently prone to forming congested zones resembling traffic jams. Although such queuing phenomena might not be present in physiological conditions, it has been proposed that they can arise from a variety of causes, including random pausing of RNAPs along the DNA~\cite{Klumpp2008} as well as ribosome stalling induced by translation-inhibiting antibiotics~\cite{Kavcic2020}.

To capture the collective dynamics of ribosomes and RNA polymerases moving along linear templates without overtaking, MacDonald and coworkers introduced the Totally Asymmetric Simple Exclusion Process (TASEP), which became the paradigmatic example of a driven lattice gas model~\cite{MacDonald1968}. In its basic formulation, the TASEP describes particles hopping unidirectionally along a one-dimensional lattice with simple exclusion: each site can be occupied by at most one particle. Although this model was initially proposed to describe ribosome traffic along mRNA during translation, it has since been adapted to transcription and other biological transport processes, see e.g.~\cite{Blythe2007, Klumpp2008, tripathi2008interacting, Wang_minimal_2014, van_crowding_2017, mines_slow_2022}.
Over time, extensions of the TASEP have been developed to better capture biological complexity, such as incorporating finite particle size~\cite{shaw_totally_2003}, internal stepping cycles~\cite{tripathi2008interacting, klumpp_effects_2008, ciandrini_role_2010, Wang_minimal_2014, Rousset2019}, site-dependent hopping~\cite{shaw_mean_2004, szavits2018deciphering, erdmann2020key, ciandrini_tasepy_2023}, and obstacle-induced stochastic pausing dynamics~\cite{turci_transport_2013, waclaw_totally_2019}. In particular, TASEP models that include random pausing and unpausing events have been proposed to describe RNA polymerase behavior during transcription, where RNAPs can pause stochastically along the DNA~\cite{klumpp_effects_2008, Wang_minimal_2014}
or be hindered by bound regulatory proteins~\cite{van_crowding_2017, mines_slow_2022, zhu_transcription_2022}. 
However, it is important to recognize that the TASEP remains an idealized model: most theoretical studies assume infinite system sizes, rely on mean-field approximations, and focus on regimes where bulk properties dominate system behavior. In contrast, biological systems operate with finite lattice lengths, specific initiation and elongation rates, and well-defined pausing dynamics. It is not always easy to assess beforehand to which extent such modeling assumptions will be respected.\\
In this article, we focus on a framework that takes into account stochastic particle pausing to study the translation process under sub-lethal antibiotic stress, where ribosomes can transiently stall due to antibiotic binding. Despite extensive work on many TASEP variants, the question how stochastic pausing shapes translation under biologically relevant conditions remains poorly understood. Here, we bridge this gap by a systematic study of a pausing TASEP model, constrained by measured biological parameters.
We develop a stochastic TASEP model that explicitly incorporates ribosome pausing and unpausing dynamics driven by antibiotic binding and unbinding, using biologically relevant parameters. To render the complex dynamics analytically tractable while still preserving biological relevance, we introduce a Single-Cluster approximation that captures the system behavior across experimentally measured initiation, elongation, and pausing parameters. Within this framework, we derive analytical expressions for key observables, including the ribosome density, the ribosomal current (protein synthesis rate), as well as the fractions of \textit{paused} (immobilized by antibiotic binding), \textit{jammed} (blocked by exclusion), and actively translating ribosomes. Our theoretical predictions show excellent agreement with extensive stochastic simulations. The model reveals how collective ribosome dynamics, strongly modulated by initiation rates and gene length, govern the response to translation-inhibiting antibiotics, an effect largely overlooked in previous theoretical approaches.
Another key feature, often absent in previous models, is that antibiotic unbinding times are comparable to mRNA degradation timescales, requiring explicit consideration of finite mRNA lifetimes when assessing translation dynamics under stress. To address this, we further extend our model to incorporate transcript degradation, thus allowing us to capture the interplay between ribosome traffic, stochastic pausing, and mRNA stability.
Within this biologically constrained framework, our model predicts that, excluding other potential regulatory feedbacks, inhibition of translation by antibiotics  becomes strongly dependent on gene length, with longer mRNAs more susceptible to disruption. 
We validate our predictions against experimental measurements of short and long gene expression under chloramphenicol treatment~\cite{zhang_decrease_2020}, finding that our analytical model quantitatively reproduces the observed trends and highlights the critical role of collective ribosome dynamics in mediating antibiotic action.
\section{Biological Context and Model}
\label{sec:bio_model}

Translation is one of the most common and critical cellular processes targeted by antibiotics. These compounds interfere with protein synthesis through a variety of mechanisms, including disrupting ribosome assembly, blocking initiation or recycling steps, interfering with the recruitment of charged tRNAs or translation factors, and inducing mistranslation~\cite{Kavcic2020,Walsh2004}. Among these modes of action, a biologically significant class of antibiotics --such as chloramphenicol, tetracycline, and erythromycin-- acts by binding to elongating ribosomes thus causing them to transiently pause on the mRNA. In this work, we focus specifically on this elongation-inhibiting mechanism. 
\subsection{The exclusion process in the context of ribosome-targeting antibiotics}

In the presence of an intracellular concentration $c$ of a ribosome-targeting antibiotic, such as chloramphenicol, antibiotic molecules can transiently interact with ribosomes at a rate $k_p$, resulting in temporary pauses of ribosomal movement along the mRNA strand. The rate at which ribosomes are hit by antibiotic molecules depends on the antibiotic concentration and the binding rate constant $k_{\text{on}}$ through the relation

\begin{equation}
    k_p = k_{\text{on}} c.
\end{equation}
Paused ribosomes then resume elongation with an `unpausing' rate \( k_u \), which depends on the antibiotic-ribosome unbinding. For chloramphenicol, the rate constant $k_\text{on}$ has been measured to be approximately \( 5.6 \times 10^{-4} \, \mu M^{-1} s^{-1} \), and the unpausing rate $k_u$ to be \( 1.4 \times 10^{-3} \, s^{-1} \)~\cite{harvey_how_1980}. Even though measuring the binding and unbinding rates of antibiotics is difficult, it has been shown that antibiotics like tetracycline~\cite{Chopra_1978, berens_tetracycline-binding_2001} and erythromycin~\cite{Pestka1974} exhibit unbinding times similar to chloramphenicol, while binding rate constants can vary significantly. 

\begin{figure}[ht]
    \centering
    \includegraphics[width=\linewidth]{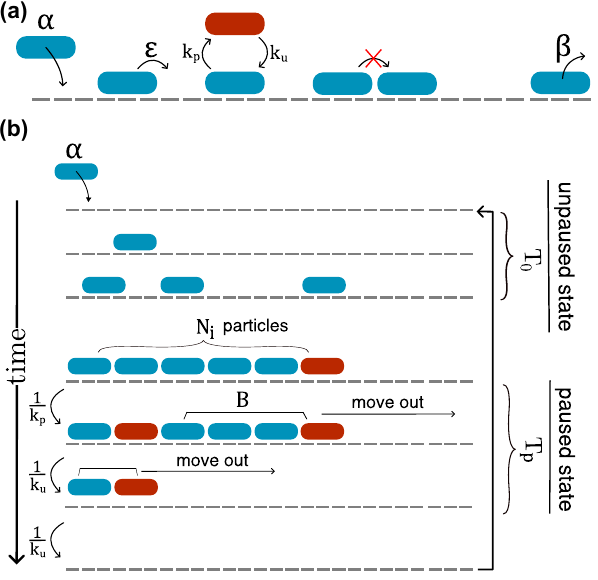}
    \caption{(a) Schematic of the TASEP where each particle (ribosome) covers \(\ell\) sites (codons), here \(\ell = 3\), for illustration. Particles enter (ribosomes initiate translation) at a rate \(\alpha\), advance (elongate) at a rate \(\epsilon\) provided the next site (codon) is unoccupied, pause and unpause with rates \(k_p\) and \(k_u\), respectively, and exit (terminate translation) at rate \(\beta\). (b) Transition from the unpaused state (zero paused particles on the lattice) to the paused state. Initially, particles progressively fill the lattice until a pausing event occurs at a random position \(X_f\), creating an initial cluster of size \( N_i = X_f/\ell\) particles. This assumes the filling of the lattice happens fast compared to the unpausing rate $k_u$. Subsequently the cluster dissolves, with discrete (average) batches of \(B\) particles each released by an unpausing event.
    }
    \label{fig:schematic_setup}
\end{figure}

To model this class of translation-inhibiting antibiotics, we develop an exclusion process in which particles (ribosomes) undergo stochastic pausing and unpausing as they progress along a one-dimensional lattice (the mRNA). This captures the essential features of ribosome dynamics under sub-lethal antibiotic stress.
Beyond pausing dynamics, translation is governed by three primary stochastic rates of the standard TASEP: the initiation rate \( \alpha \) rate, at which ribosomes bind the mRNA entry site; the elongation rate \( \epsilon \) at which ribosomes move codon by codon, provided that the next codon is empty; and the termination rate \( \beta \) characterising the detachment of the ribosomes from the last site. In the exclusion process, the mRNA strand is represented as a unidimensional discrete lattice, where each site stands for a codon. This is sketched in Fig.~\ref{fig:schematic_setup}(a). 
When speaking about the model we will refer to a ribosome as a \textit{particle}, to an mRNA strand as \textit{lattice}, and to the initiation/termination rate as \textit{entry} and \textit{exit rate}. 

Two quantities are important, both in terms of modelling and in terms of observation: density and current. The (space-averaged) ribosome density $\rho$ is the number of ribosomes present on the lattice at a given moment, divided by the number of codons constituting the lattice. The local variant of this quantity, for a single codon site, may be interpreted as the probability of finding a ribosome on this specific site. Finally, the local current is the number of ribosomes crossing any specific codon per unit of time. At the last site, the exit current therefore corresponds to the protein synthesis rate. 

In {\it E. coli}, the elongation rate typically lies between $10$ and $20$ codons per second, while the initiation rate is estimated between \( 0.1 \) and \( 1 \) ribosomes per second~\cite{Dai2016}. The termination process is generally considered fast relative to other translation events~\cite{Ciandrini2013}. Ribosomes typically occupy a footprint of $\ell = 10$ codons \cite{milo2015cell}, and in {\it E. coli} the average mRNA length is $255$ codons with a total range between $L = 15 - 2368$ codons~\cite{balakrishnan_principles_2022, milo2015cell}.

\subsection{Single-cluster picture}
\label{sec:single_cluster}
The full particle dynamics with pausing/unpausing due to the action of antibiotics is a very complex problem. However, a simplified picture emerges after observing that the unbinding rates of antibiotics such as chloramphenicol are typically three to four orders of magnitude lower than the ribosomal elongation rate (see values given above). Therefore the antibiotic remains bound to a ribosome for a time equivalent to many elongation steps. The pronounced separation of timescales would therefore suggest a simplified representation in which a single paused ribosome initiates a traffic jam, propagating backwards to rapidly reach a single, fully congested cluster reaching back to the initiation site, Fig.~\ref{fig:schematic_setup}(b). We refer to this as the \emph{Single-Cluster Approximation}, which allows for a simplified yet tractable analytical description of the system's behavior in biologically relevant regimes.

As a simple back-of-the-envelope calculation backing up this picture, we take the reported rate binding constant \(k_{on}\) for chloramphenicol~\cite{harvey_how_1980} at a concentration of \(10\,\mu\text{M}\). Under these conditions, the average time for a ribosome to get hit by the antibiotic is approximately \(180\,\text{s}\). Given that a typical ribosome elongation step occurs at \(\sim 0.1\,\text{s}\) per codon (i.e.,  \(10\,\text{aa/s}\)~\cite{Dai2016}), a single ribosome would traverse \(\sim 1800\) codons before experiencing a binding event. Since the average protein length is around \(255\) codons~\cite{balakrishnan_principles_2022}, multiple binding events on the same ribosome are thus highly unlikely.
Additionally, the unbinding time for chloramphenicol is around \(\sim 700\,\text{s}\)~\cite{harvey_how_1980}. With a time interval between initiation events of \(\sim 10\,\text{s}\), there can be roughly 70 initiation events on an mRNA before the antibiotic unbinds from a ribosome paused on it. Assuming each ribosome footprint extends across \(10\) codons, this leads to a cluster size of \(\sim 700\) codons, which far exceeds the typical protein length~\cite{balakrishnan_principles_2022}. These estimates support the conclusion that a \emph{Single-Cluster approximation} is biologically reasonable under these conditions.
This simplified picture entails simplified dynamics, sketched in Fig.~\ref{fig:schematic_setup}(b), which are as follows. During the first phase in the single cluster approximation, particles enter at the initiation site (with entry rate $\alpha$), and these particles move across the lattice following the standard TASEP dynamics. As soon as a first pausing event occurs at some random position, particles downstream from this point evacuate and terminate. Simultaneously, further particles enter until the upstream section to the paused particle is fully filled. Neglecting the short-lived transient phases, we are therefore looking a system switching between two main states: a state with zero paused particle, obeying standard TASEP dynamics, which we refer to as the \emph{unpaused} state, by opposition to its counterpart, the \emph{paused state}, which has at least one paused particle on the lattice, entailing a jam of particles.
Once the paused state is established, all particles may undergo pausing or unpausing events. This establishes an equilibrium of paused and unpaused particles, thereby splitting the jam into batches of $B$ particles separated by paused particles.
We assume that whenever the leading paused particle unpauses, the next batch of particles can rapidly exit the lattice in a short-lived transient phase, which we do not intend to resolve. We assume, in this picture, that the batch that is set free leaves the lattice instantaneously. 
The system exits the paused state only once all paused ribosomes have resumed translation and completely exited the lattice, ensuring that the entire sequence of discrete ribosome batches has left the system.
In a finite cluster of size \(N\), the mean batch‐size is
\begin{equation}\label{eq:block_of_particles_finite}
B = \frac{k_u}{k_p}\Bigl[1 - \bigl(\tfrac{k_p}{k_p + k_u}\bigr)^{N}\Bigr] + 1,
\end{equation}
where, the “+1” accounts for the leading paused ribosome. This result follows (see Appendix A) by computing the expected number of consecutive unpaused particles before a pause under a geometric law truncated at \(N\)~\cite{keisers2024finite}, and it reduces smoothly to the infinite‐cluster result \(B = k_u/k_p + 1\) as \(N\to\infty\). Note that, despite a formal divergence of $B$ in the limit \(k_p \to 0\), this is not an issue: in this regime, pausing events become vanishingly rare, and thus the contribution from the paused system discussed here becomes irrelevant.

After a timespan \(T_p\), the lattice is fully emptied, and the entire process restarts with particles entering an empty lattice. 
 
It is worth summarizing the conditions for this picture to be a reasonable approximation:
(i) pausing events must be rare compared to elongation steps ($k_p \ll \epsilon$), (ii) termination cannot limit the dynamics ($\beta \approx \epsilon$ is a safe condition) (iii) the segment cannot be too long, in a sense to be refined, since otherwise there would be formation of additional blockages as the paused state dissolves, which would invalidate the simplification of a single cluster.
As estimated above, these conditions are met at least for some classes of antibiotics.

\section{Results: analysis of the model}
We now need to better characterize the dynamics of these main states, which essentially amounts to understanding the size of the duration of the paused state, as well as the process by which the system returns to an unpaused state.

\subsection{Unpaused State}\label{subsection:unpaused state}
To characterize the unpaused state (zero paused particles on the lattice, thus labeled with the suffix $0$), we examine the evolution from an initially empty lattice towards a steady-state density profile, starting when particles enter with rate \( \alpha \) and move unidirectionally with an elongation rate \( \epsilon \). In the absence of pausing, the system follows the standard TASEP with extended particles~\cite{shaw_totally_2003}. As particles enter the empty lattice, a density front --often referred to as a \emph{domain wall}-- emerges, separating a high-density region behind the wall from a zero-density region ahead of it \cite{kolomeisky_phase_1998, santen_asymmetric_2002, cividini_exact_2014}.
The domain wall propagates through the lattice with velocity
\begin{equation}
    v = \epsilon - \alpha.
\end{equation}
The position of the domain wall or shock, $x_s$, at time \( t \) follows as
\begin{equation}\label{eq:domain_wall_pos}
    x_s(t) = v t.
\end{equation}
The time required for the domain wall to traverse a lattice of length \( L \), denoted as \( \Tss \), thus is 
\begin{equation}\label{eq:time_until_steady_state}
    \Tss = \frac{L}{\epsilon - \alpha}.
\end{equation}

After this time, the lattice reaches a steady state characterized by a constant particle density and current~\cite{szavits-nossan_dynamics_2020}. Under standard TASEP assumptions for extended particles, these quantities are given by
\begin{equation}\label{eq:current_standard}
    J_0 = \frac{\alpha(\epsilon - \alpha)}{\epsilon + \alpha (\ell -1)}, \quad
    \rho_0 = \frac{\alpha}{\epsilon + \alpha(\ell -1)},
\end{equation}
where \(J_0\) represents the steady-state particle current, and \(\rho_0\) is the corresponding particle density, which follows directly from mean-field approximations commonly applied in exclusion processes~\cite{shaw_totally_2003}. Here, $\ell$ is the particle footprint in terms of lattice sites.
\subsubsection{Lifetime Distribution of the Unpaused State}
The random time $T_0$ until the first pausing event can be analysed by viewing pauses as a \emph{simple point process}, in which times at which pauses occur form a random set of points on the time axis.
Particles pause independently following a stochastic Poisson process with rate $k_p$.
However, the number of particles in the lattice 
grows linearly in time with the injection current $ J_0 $, until the steady state time $ \Tss $,
after which it remains constant:
\begin{equation}
    N =  
    \begin{cases}
    J_0 t
    , & t  < \Tss\\[6pt]
    \rho_0 L 
    , & t  \geq \Tss \,.
    \end{cases} 
\end{equation}
Therefore we are dealing with an \emph{inhomogeneous Poisson process}~\cite{kingman1993poisson} (also called a non‑stationary Poisson process) for the first phase, $t<\Tss$. Here, the rate at which pausing occurs anywhere in the system is thus time-dependent, and given as $\lambda(t) \;=\; k_p\,N(t)$.
Once the steady state has been reached, during the second phase ($t\ge \Tss)$, this rate saturates as $N$ saturates. All in all, we thus have
the rate for the system to transition from an unpaused to a paused state as
\begin{equation}\label{eq:instant_pausing_rate}
    \lambda(t) =
    \begin{cases}
        J_0 k_p t , & t  < \Tss\\[6pt]
        \rho_0 L k_p , & t  \geq \Tss \,.
    \end{cases} 
\end{equation}

The probability that no pausing has occurred up to time $t$ is the survival function of the unpaused state,  a standard result for Poisson processes~\cite{ross1996stochastic}:
\begin{equation}
    S(t)\;=\;\exp\!\bigl[-\Lambda(t)\bigr].
    \label{eq:survival}
\end{equation}
Here $\Lambda(t)$ is the \emph{cumulative} (or integrated) intensity, i.e the expected number of pausing events up to time $t$, given by
\begin{equation}\label{eq:survival_prob_unclogged_state}
    \Lambda(t) = \int_0^{t} \lambda(s)\,ds =
    \begin{cases}
        a\,t^2, & t < \Tss, \\[6pt]
        a\,\Tss^2 + b\,(t - \Tss), & t \geq \Tss,
    \end{cases}
\end{equation}
where we define \(a := J_0 k_p/2\) and \(b := \rho_0 L k_p\).

We now define the probability density $F(t)$ for the first pausing event, anywhere in the system,
as $F(t) = \lambda(t)\,S(t)$: for a first pausing event  to take place in the time interval $dt$ is the product of the survival probability $S(t)$, ensuring the system has not paused before $t$, and the probability $\lambda(t) \, dt$, for pausing to take place in the following interval $dt$.  Thus
\begin{equation}\label{eq:PDF_time_till_pause}
    F(t) = \lambda(t)\,S(t) =
    \begin{cases}
        2a\, e^{-a t^2}, & t < \Tss, \\[6pt]
        b\, e^{-a \Tss^2 - b (t - \Tss)}, & t \geq \Tss.
    \end{cases}
\end{equation}
Since $\Lambda(t)$ is not constant, this function is Rayleigh-like up to $\Tss$, reflecting the quadratic growth of $\Lambda(t)$, while for $t \ge \Tss$ it acquires the familiar
exponential tail of a homogeneous Poisson process.

\subsubsection{Expected Lifetime of the Unpaused State}
The expected duration of the unpaused state, denoted as \( \Tfp \), now follows as the average time when the first pausing event occurs, thus using the function $F(t)$ in Eq.~\eqref{eq:PDF_time_till_pause} to weight the average:
\begin{equation*}
    \Tfp = \int^{\Tss}_0 t \, F(t)\, dt
\end{equation*}
Note that $F(t)$ is the probability density function for the distribution of the first pausing time. 
Splitting up the integral as
\begin{equation}\label{eq:time_in_zero_state}
    \Tfp = \int^{\Tss}_0 t \, F(t)\, dt + \int^\infty_{\Tss} t \, F(t)\, dt 
\end{equation} 
identifies two contributions, where the first integral corresponds to the expected duration of the unpaused state during the period characterized by an inhomogeneous Poisson process, described mathematically by a truncated Rayleigh distribution~\cite{kingman1993poisson}. Evaluating this integral explicitly yields
\begin{equation}
     \int^{\Tss}_0 t\, F(t)\, dt = \frac{\sqrt{\pi}}{2\sqrt{a}}\, \operatorname{erf}(\sqrt{a}\,\Tss) - \Tss e^{-a\Tss^2},
     \label{eq:T_0a}
\end{equation}
with \(\operatorname{erf}\) denoting the error function.% and \(a := J_0 k_p/2\).

The second contribution to
Eq.~\eqref{eq:time_in_zero_state} from first unpaused events taking place after the system has reached steady state, {\it i.e.} after the domain  wall has arrived at the exit site. The lifetime decays exponentially and can be described by
\begin{equation}
    \int^\infty_{\Tss} t \, F(t)\, dt = \frac{e^{-b \Tss}}{b} + \Tss e^{-a\Tss^2} \,.
    \label{eq:T_0b}
\end{equation}
Adding Eqs.~(\ref{eq:T_0a}) and (\ref{eq:T_0b}) we obtain 
\begin{equation}\label{eq:expected_time_until_pause}
    \Tfp = \frac{\sqrt{\pi}}{2\sqrt{a}}\, \operatorname{erf}(\sqrt{a}\,\Tss) + \frac{e^{-b \Tss}}{b}.
\end{equation}

\begin{figure}
    \centering
    \includegraphics[width=\columnwidth]{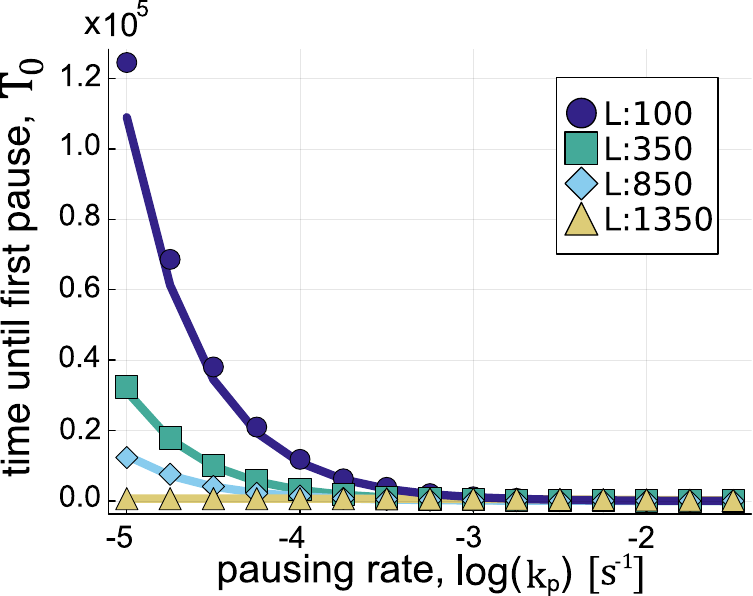}
    \caption{Lifetime of the unpaused state, {\it i.e.} expected time until the first pausing event, as a function of the pausing rate, for different lattice lengths $L$. Analytical results (lines) closely match simulation data (markers). The simulation parameters here are $\alpha = 0.2$, $\epsilon = 20$, $k_u = 1.4 \times 10^{-3} $}
    \label{fig:time_until_first_pause}
\end{figure}

As illustrated in Fig.~\ref{fig:time_until_first_pause}, this analytical result captures the dynamics very accurately. 

\subsection{Paused state} 

We now characterise the paused state by the average length of the jam it entails, as well as its lifetime.

\subsubsection{Entering the paused state and expected first pausing position}

In order to quantify the length of the jam in the paused state we must identify the position at which the first pausing event occurs. To this end we first consider the instantaneous position of the domain wall, \( x_s(t) \), as defined in Eq.~\eqref{eq:domain_wall_pos}. Any particle behind this domain wall has an equal probability of pausing. Because the particle density upstream of the density front is homogeneous, pausing is equally likely to occur on each upstream site. Consequently, as long as the time of interest is \( t < \Tss \), the expected position of the first pausing event, \(\xfp(t)\), is at half the domain wall position, \( x_s(t)/2 \). However, once the domain wall reaches the exit site, at \( t \geq \Tss \), the density is uniform across the entire lattice, and thus the expected pausing position is simply the midpoint \( L/2 \). Therefore the position \(\xfp(t) \) at which, on average, the first pausing event is expected to occur can be summarized as
\begin{equation}\label{eq:pausing_position_xfp}
    \xfp(t) = 
    \begin{cases}
        \frac{v t}{2}, & t < \Tss \\[6pt]
        \frac{L}{2}, & t \geq \Tss
    \end{cases}
\end{equation} 
To calculate the average position \( X_f\) where the paused state is initiated, by the first pausing event, we weigh the  (time-dependent) position $x_f$ where pausing is expected to occur by the probability density function \(F(t)\) from Eq.~\eqref{eq:PDF_time_till_pause}, and integrate over all possible times leading to
\begin{equation}\label{eq:first_pausing_position}
    X_f = \int^{\Tss}_0 \frac{v t}{2} F(t)\, dt + \int^\infty_{\Tss} \frac{L}{2} F(t)\, dt.
\end{equation}
Evaluating this integral explicitly, we find
\begin{equation}\label{eq:pausing_position}
    X_f = \frac{v}{2}\left[\frac{\sqrt{\pi}}{2\sqrt{a}} \operatorname{erf}(\sqrt{a}\,\Tss) - \Tss e^{-a \Tss^{2}}\right] + \frac{L}{2} e^{-a\Tss^2}.
\end{equation}

\begin{figure}
    \centering
    \includegraphics[width=\columnwidth]{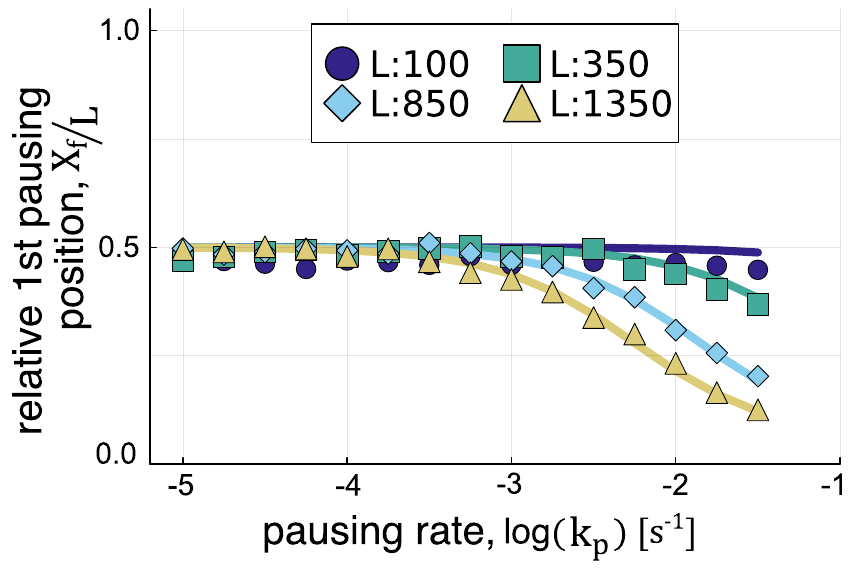}
    \caption{Relative position ($X_f/L$) of the expected first pausing event for various lattice lengths as a function of the pausing rate $k_p$. The relative pausing position shifts towards the lattice midpoint \( 0.5 \) as lattice length increases or pausing rate decreases. Analytical results (lines) closely match the simulation (markers). The parameters are the same as in Fig.~\ref{fig:time_until_first_pause}.}
    \label{fig:relative_pausing_position}
\end{figure}

After the first pausing event, expected to occur at position $X_f$ at time \( \Tfp \), the system transitions into the \emph{paused state}. 
From Eq.(\ref{eq:pausing_position}) it now becomes clear that the lattice size plays an important role: for a given pausing rate, the larger the lattice $L$, the further ahead of the mid-position $L/2$ the first pausing will occur. This is illustrated and validated by simulation results in Fig.~\ref{fig:relative_pausing_position}. 

\subsubsection{Lifetime of the paused state}
To compute the lifetime of the paused state we need to evaluate the time to dissolve it by successive release of discrete batches of particles (see Fig.~\ref{fig:schematic_setup}). The number of particles in the initial cluster is, on average, \( \CNi := X_f/\ell\) when considering the filling of the initial cluster as a fast process. This means that the loading of all the particles in the cluster must be faster than the unpausing time, which is roughly translated in the condition $ N_\text{i} / \alpha \gg 1/ k_u$. The assumptions of the single-cluster approximation are discussed in Sec.~\ref{sec:single_cluster}.

In order to evaluate the time $T_p$ spent in the paused state we need to compute the expected number of successive unpausing events $n_B$  required to empty the lattice, thus returning to an unpaused state. To do this we determine the number $n_B$ of particle batches —each of size \( B \), defined by Eq.~\eqref{eq:block_of_particles_finite} --which must sequentially exit to dissolve the initial cluster. Thus, the average number of required unpausing events \( n_B \) to dissolve an initial cluster with \( \CNi \) particles by batches of $B$ particles can be approximated by
\begin{equation}
    n_B = \left\lceil \frac{\CNi}{B} \right\rceil, 
    \label{eq:expected_n_B}
\end{equation}
where the ceiling $\lceil z \rceil$ returns the least integer greater than or equal to $z$.

Each batch of particles requires an average time \( 1/k_u \) for its leading paused particle to become unpaused, after which it rapidly leaves the lattice. Consequently, we can express the total time required to dissolve the entire cluster, \( T_p \), and thus the lifetime of the paused state, as 
\begin{equation}
    T_p = \frac{n_B}{k_u}.
\end{equation}

\begin{figure}
    \centering
    \includegraphics[width=\columnwidth]{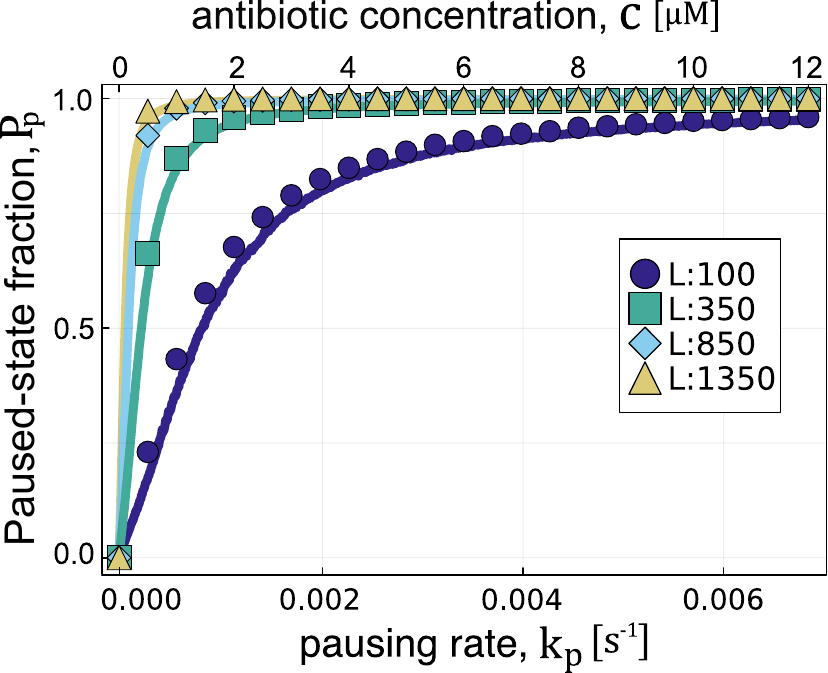}
    \caption{Fraction of time spent in the paused state for different lattice lengths. The probability of being in the unpaused state can be obtained as $1 - P_p$. Symbols are simulation data, lines are predictions within the Single-Cluster approximation.}
    \label{fig:weighted_and_unimpeded_translation}
\end{figure}

Considering the expected time spent in the unpaused state \( \Tfp \), given by Eq.~\eqref{eq:time_in_zero_state}, we can quantify the fractions of time that the system spends either in the paused state (\( P_p \)) or in the unpaused state (\( P_0 \)) as follows: 
\begin{equation}\label{eq:P0 and PC}
    P_0 = \frac{\Tfp}{\Tfp + T_p}, \quad\text{and}\quad  P_p = \frac{T_p}{\Tfp + T_p}.
\end{equation}
These analytical predictions closely match simulation results, as illustrated in Fig.~\ref{fig:weighted_and_unimpeded_translation}.

The lattice length influences these fractions via two mechanisms, which both favor the paused state. 
First, a larger lattice results in a higher total particle number, thereby shortening the expected duration of the unpaused state (see Fig.~\ref{fig:time_until_first_pause}).
Second, a larger lattice leads to larger initial clusters which thus require more time to dissolve, consequently also increasing the lifetime of the paused state. 

\subsection{Average observable quantities}
Having established the fraction of time the system spends in the paused and unpaused states, respectively, we can now calculate the weighted averages of observable quantities.

\subsubsection{Particle density}\label{sec:particle_density}

The total particle density, \( \rho \), is obtained as a weighted average of the paused and unpaused states,
\begin{equation}\label{eq:total_density}
    \rho = P_0 \rho_0 + P_p \rho_p,
\end{equation}
where the probabilities \( P_0 \) and \( P_p \) are defined in Eq.~\eqref{eq:P0 and PC}, and the steady-state density in the unpaused state \( \rho_0 \) is given by the standard TASEP expression, Eq.~\eqref{eq:current_standard}.

To analytically determine the density  in the paused state, \(\rho_p\), we first calculate the average number of particles during the dissolution process for a given initial cluster length \(\CNi\), and subsequently average over all possible initial cluster lengths.

As described previously, clusters dissolve sequentially in discrete batches of \(B\) particles, each requiring an average time \(1/k_u\) to detach from the main cluster and leave the lattice in a transient assumed to be fast. For simplicity, when solving the equations below, we assume the limit of large $N$ in Eq.~\eqref{eq:block_of_particles_finite}, such that \( B = k_u/k_p + 1 \). This approximation allows us to analytically solve the equations without introducing significant errors (see Appendix~\ref{sec:appendix_batch}).
Thus, the time-averaged number of particles in the cluster $\langle N_c \rangle$, for a given initial cluster size \(\CNi\), is calculated by averaging the particle number over each dissolution step until the lattice is empty:
\begin{equation}\label{eq:density(N)}
    \langle N_c \rangle= \frac{1}{n_B} \sum_{q=0}^{n_B-1}\left[\CNi - B q\right],
\end{equation}
where \(n_B\), defined in Eq.~\eqref{eq:expected_n_B}, represents the total number of unpausing events required to fully dissolve the cluster. We note that $n_B$ is a function of the initial size of the cluster $\CNi$, but above we omit this dependency here for clarity. The summation in Eq.~\eqref{eq:density(N)} ends at \(q=n_B-1\) because the lattice becomes empty at \(q=n_B\).
To determine the average particle density in the paused state (\(\rho_p\)), we consider all possible initial cluster sizes \(\CNi = 1, \dots, \Cmax\), where \(\Cmax := x_s / \ell\) represents the largest possible number of particles that can form a cluster before the first pausing event occurs---i.e., when the leading particle (located at the domain wall position \(x_s\)) becomes paused and initiates the blockage.
Because larger initial clusters persist longer, and therefore contribute more significantly to the overall density, we weight each cluster size by its corresponding dissolution time, proportional to the number of required unpausing events \(n_B(\CNi)\). The average density in the paused state is thus computed as the weighted average:
\begin{equation}\label{eq:weighted_density_clogged}
    \rho_p = \frac{\displaystyle\sum^{\Cmax}_{\CNi=1} n_B(\CNi) \, \cfrac{\langle N_c \rangle}{L}}{\displaystyle\sum_{\CNi=1}^{\Cmax} n_B(\CNi)}.
\end{equation}
Evaluating this expression explicitly yields
\begin{equation}
    \rho_p = \frac{2 \Cmax + 3B + 1}{6L},
\end{equation}
which shows how particle density depends on the maximum initial cluster size \(\Cmax\), the batch size \(B\), and the lattice length \(L\).

Having determined the density of both unpaused and paused states (\(\rho_0\) and \(\rho_p\)) and their respective probabilities, Eq.~\eqref{eq:total_density} can now be used to fully characterize the system density as a function of the  pausing rate \(k_p\).

\begin{figure}[h]
    \centering
    \includegraphics[width=\columnwidth]{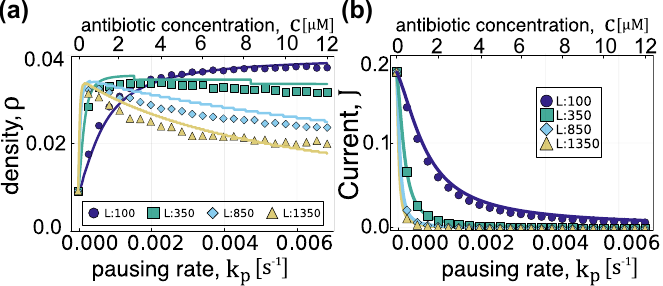}%
    \caption{(a) Particle density as a function of the pausing rate \( k_p \), for different lattice sizes. Noteably, the overall density exhibit a non-monotonic behavior due to the first pausing position being moved closer to the initiation site for higher pausing rates. (b) The current decreases monotonically with the pausing rate. Note that shorter lattices exhibit a higher current compared to longer ones for the same pausing rate. Here, $\alpha$, $k_u$ and $\epsilon$ are equal to $0.2$, $0.0014$ and $20$ respectively.}
    \label{fig:density_and_current}
\end{figure}

Figure~\ref{fig:density_and_current}(a) illustrates this dependency. Initially, increasing \( k_p \) results in a higher likelihood of entering the paused state, which leads to an increase in particle density. However, this trend saturates beyond a crossover value for \( k_p \), as the system spends most of its time fully paused. Further increasing \( k_p \) then shifts the position of the initial paused particle closer to the initiation site, effectively reducing the initial cluster size and consequently lowering the overall density. This non-monotonic behavior is observed in Fig.~\ref{fig:density_and_current}a).

\subsubsection{Current}
Similarly to the density, we can also compute the average current, $J$, as a weighted sum:
\begin{equation}\label{eq:total_current}
    J = P_0 \, J_0 + P_p \, J_p.
\end{equation}
The contribution from the unpaused state is given simply by the zero-paused particle current \( J_0 \), as described by the standard TASEP expression for extended particles, Eq.~\eqref{eq:current_standard}. Conversely, the current contribution \( J_p \) from the paused phase is more complex.
It depends on the (average) number $N_\text{i}$ of particles in the first cluster after the first pausing event, and on the time $T_p$ required for this cluster to dissolve. Explicitly, the contributions from the paused and unpaused states are
\begin{equation}%[floatfix]
    J_0 = \frac{\alpha(\epsilon - \alpha)}{\epsilon + \alpha (\ell -1)}.
    \qquad\mbox{and}\qquad 
    J_p = \frac{N_\text{i}}{T_p} \,.
\end{equation}
Figure~\ref{fig:density_and_current}b) shows that this analytical solution agrees well with the simulation. In contrast to the density, the current is monotonic for all lattice sizes: increasing the pausing rate will always reduce the current. 

\subsubsection{Microscopic particle states}
Having characterized the global properties of the system in terms of density and current, we now turn our attention to the microscopic dynamics by categorizing the states of individual particles.

In order to do that, we divide particles in three distinct categories: paused, mobile, and jammed (not paused, but blocked by exclusion by other particles downstream).

The \emph{fraction of paused fraction}, \( f_p \), represents the fraction of particles that are paused. Its equilibrium value is determined from the pausing and unpausing rates as
\begin{equation}\label{eq:paused_fraction}
    f_p = \frac{k_p}{k_u + k_p}.
\end{equation}

Another relevant quantity is the \emph{mobile fraction}, \( f_m \), which accounts for particles that can move to the next site. Only this group of particles contributes to the overall current. 
The current can be expressed in terms of the number of mobile particles as $J = \epsilon \, N_m / L$, this leads to $f_m = N_m / N = (J \, L / \epsilon)/N$, and thus this fraction can be expressed as
\begin{equation}
    f_m = \frac{J}{\epsilon \, \rho}.
\end{equation}

Since every particle must belong to one of the three categories (paused, jammed, or mobile), their fractions sum to one.
Thus, the \emph{jammed fraction}, \( f_j \) --which represents particles that are not paused but are blocked due to steric hindrance-- can be defined by
\begin{equation}\label{eq:jammed_fraction}
    f_j = 1 - (f_m + f_p).
\end{equation}

In the next section we will describe how neglecting the jammed fraction leads to overestimate the actively translating ribosomes under inhibition of translation by ribosome-targeting antibiotics.

\section{Results: biological applications}

In this section we discuss several biological situation which have motivated us to develop the model of a TASEP with stochastically pausing and unpausing particles.

\subsection{The model predicts antibiotics-induced ribosome jamming and sequestration }

In models of quantitative physiology a key quantities is the biomass rate production (growth rate), which is often identified with protein synthesis made by ribosomes, i.e. the current $J$ in this model multiplied by the entire set of mRNAs.
Thus, in phenomenologycal models of quantitative physiology, ``active'' ribosomes at different environmental conditions determine biomass production and cellular growth~\cite{Dai2016, balakrishnan_principles_2022}. 
In the literature, the definition of an active ribosome varies. Some studies define active ribosomes as those contributing to the current~\cite{Dai2016, balakrishnan_principles_2022}, while others classify any ribosome not in a paused state as active~\cite{Wang_minimal_2014}
, regardless of whether they are temporarily blocked by downstream particles. Moreover, common models in quantitative physiology which link protein synthesis and mass growth neglect the role of ribosome interference (traffic). Here, we propose a more precise definition of active particles based on the previously introduced groups of paused, jammed, and mobile particles.

This allows us to compare the impact of elongation-targeting antibiotics (like chloramphenicol) on the population of translating ribosomes.
\begin{figure}[h]
    \centering
    \includegraphics[width=\columnwidth]{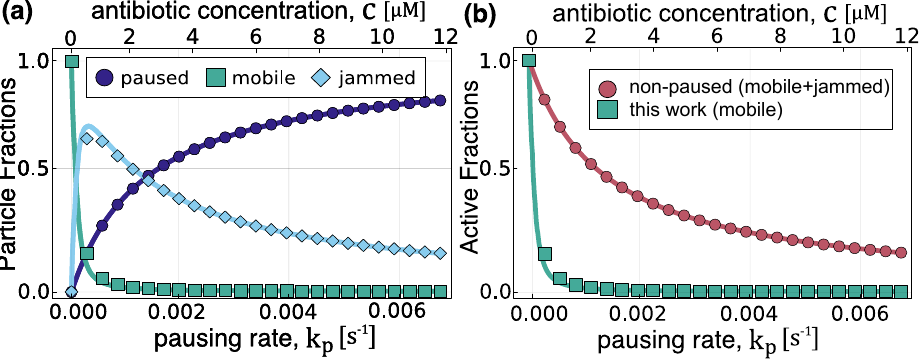}%
    \caption{(a) shows how the fractions of jammed, mobile, and paused particles evolve with different pausing rates. (b) compares various definitions of active particles (those contributing to the current) as a function of the pausing rate $k_p$. Steric interactions are often neglected when estimating the fraction of active ribosomes, so that jammed particles are still counted as contributing to the current (red circles). In contrast, our approach defines active particles only as those that are mobile (green squares). In this figure, $\alpha = 0.02$, $\epsilon = 20$, and $k_u = 1.4 \times 10^{-3}$.
    }
    \label{fig:particle_fractions}
\end{figure}
Figure~\ref{fig:particle_fractions}(a) shows the behavior of these fractions as functions of the pausing rate \( k_p \), demonstrating excellent agreement between analytical predictions and simulation results. Intuitively, the fraction of ribosomes impacted by antibiotics (paused particles) steadily increases with antibiotic concentration. Instead, the mobile fraction abruptly decreases even at low antibiotic concentrations. This sharp decline is explained by the concurrent rise in jammed ribosomes, which, in the absence of ribosome-interference regulatory mechanisms (e.g. mRNA degradation), would sequester the majority of ribosomes away from active translation at low antibiotic levels.

Figure~\ref{fig:particle_fractions}(b) compares active ribosomes as defined in our model (those contributing to total current, i.e., mobile particles) with the definition that neglects ribosome interference. The latter definition, typically used in the literature, considers active ribosomes as the total minus those paused, corresponding to the combined jammed and mobile fractions \( f_j + f_m \).

The model thus predicts that antibiotics have an impact in lowering the fraction of active ribosomes, not only as a direct result of the antibiotics-ribosome binding, but also because they induce jamming by ribosome interference, as suggested in~\cite{Kavcic2020}, but usually neglected in the literature. However, we highlight that this prediction does not account for potential traffic and ribosome biogenesis regulatory mechanisms~\cite{ balakrishnan_principles_2022, wu_cellular_2022}

\subsection{Collective ribosome dynamics affect hitting probability}

The \emph{hitting probability} --defined as the likelihood that a ribosome binds an antibiotic before completing translation-- is often used to estimate the impact of elongation-inhibiting antibiotics on protein synthesis. 
An expression commonly used in the literature (see, e.g.~\cite{Dai2016, zhang_decrease_2020}) is 
\begin{equation}\label{eq:P_hit_literature}
    P^{\text{lit}}_{\text{hit}} = 1 - e^{-k_p \Tss},
\end{equation}
where \( k_p \) is the antibiotic binding rate and \( \Tss \) is the time required for a ribosome to translate the full mRNA. While this expression is simple and analytically tractable, it is derived under the assumption that the mRNA is occupied by a single ribosome at a time. 

However, under physiological conditions, ribosomes are frequently loaded onto mRNAs at initiation rates that allow multiple ribosomes to elongate simultaneously on the same transcript. In this setting, each ribosome can independently pause because it is targeted by the antibiotic, so the likelihood that at least one ribosome is interrupted during translation increases with the number of ribosomes on the mRNA. This introduces a collective effect that is not captured by the single-ribosome expression in Eq.~\eqref{eq:P_hit_literature}. To incorporate this, we define a more general hitting probability that accounts for the full stochastic dynamics of ribosome loading and pausing,
\begin{equation}\label{eq:P_hit_multiple_particles}
    P_{\text{hit}} = 1 - \int_{\Tss}^{\infty} F(t)\, dt = 1 - e^{-a\,\Tss^2}, 
\end{equation}
with \(a=J_0k_p/2\). 
This relation quantifies the probability that an antibiotic molecule hits any ribosome translating the mRNA before a time $t_L$. This is equivalent to an mRNA entering the paused state before reaching the steady state, i.e. before finishing the first round of translation.

\begin{figure}[h]
    \centering
    \includegraphics[width=\columnwidth]{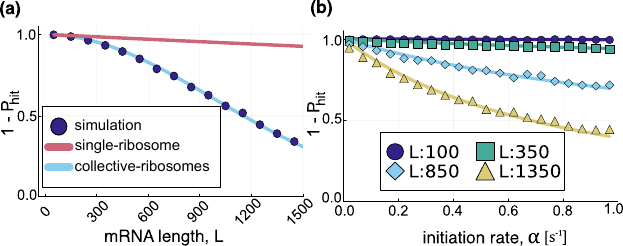}
    \caption{
    (a) Probability of completing translation without antibiotic-induced pauses for varying lattice lengths, compared to the single-ribosome prediction (red line, Eq.~\eqref{eq:P_hit_literature}). Parameters: \(\alpha = 0.25\), \(\epsilon = 20\), \(k_p = 10^{-3}\), \(k_u = 1.4\times 10^{-3}\). 
    (b) Influence of the initiation rate \(\alpha\) on the probability of unimpeded translation for different lattice lengths. Decreasing \(\alpha\) makes a translation-halting event less likely. Lines are analytical predictions; symbols show simulation data. The parameters used here are $k_p = 10^{-3}$, \(\epsilon = 20\), $k_u = 1.4\times 10^{-3}$.}
    \label{fig:hitting_probability}
\end{figure}

Figure~\ref{fig:hitting_probability}(a) illustrates the contrast between the collective hitting probability Eq.~\eqref{eq:P_hit_multiple_particles} and the single-ribosome approximation from Eq.~\eqref{eq:P_hit_literature}. For short lattices --corresponding to low ribosome numbers-- the two expressions closely agree. However, for longer lattices, where more ribosomes can simultaneously occupy the mRNA, the collective hitting probability \(P_{\text{hit}}\) increases  significantly, due to the increased likelihood that some ribosome will pause before translation completes. Figure~\ref{fig:hitting_probability}(b) further demonstrates that the probability of successful translation (i.e., no pausing before completion) decreases not only with gene length but also with the initiation rate. Here too, having more ribosomes amplifies the probability that at least one will be hit during translation.

These results highlight a non-trivial collective effect: although the chance of any single ribosome pausing at a given moment is low, the probability that at least one ribosome pauses during the time from initiation to termination, given by $t_L$, increases sharply with the number of ribosomes translating the mRNA.

The hitting probability is a useful estimate for the likelihood that translation is prematurely interrupted. In principle, to understand its impact on protein output, we should also allow for translation to resume after a pausing event. In many biological systems however, mRNAs degrade on a timescale that is comparable to or even shorter than the antibiotic unbinding time. This implies that once an mRNA is hit, translation is unlikely to recover --making the timing of the first pause a key determinant of total protein yield.

\subsection{Gene-length dependent expression differences emerge from collective dynamics}
Building on the discussion of how antibiotic stress affects translation dynamics  through pausing and collective effects, we now compare our model predictions with experimental data. One key prediction of our model is that, the longer the proteins, the more severe is the inhibition by ribosome-targeting antibiotic stress. This prediction qualitatively agrees with experimental data from Zhang {\it et al.}~\cite{zhang_decrease_2020}, who compared the expression under various antibiotic concentrations of two genes, \textit{gfpmut2} (238 codons) and \textit{lacZ} (1024 codons), integrated at identical genomic locations. They observed an increased short-to-long protein expression ratio, consistent with our prediction of length-dependent translation inhibition.

In \textit{E.~coli}, the average lifetime of an mRNA is typically between 1 and 5 minutes~\cite{balakrishnan_principles_2022}, while the unbinding time of antibiotics like chloramphenicol is much longer --around 12 minutes~\cite{harvey_how_1980}. This implies that once a ribosome is paused by an antibiotic, the mRNA is likely degraded before the ribosomes on the mRNA can resume any meaningful translation. Moreover, the pause of the ribosomes and the traffic jam generated may also trigger a cascade of signalling promoting the degradation of the affected transcripts~\cite{keisers_exclusion_2023,ikeuchi_recent_2019, Subramaniam2014}.

To connect our model to experimental measurements of protein output, we must consider the finite lifetime of mRNAs, which constrains the time available for translation. We model the distribution of mRNA degradation times as an exponential distribution,
\begin{equation}\label{eq:mRNA_expo_distr}
    \varphi_{\text{mRNA}}(t) = \frac{1}{\tau} e^{-t/\tau},
\end{equation}
where \( \varphi_{\text{mRNA}}(t) \) denotes the probability density that an mRNA is degraded at time \( t \), and \( \tau \) is the average mRNA lifetime.

In the presence of antibiotic stress, protein synthesis from a single mRNA transcript can be terminated in two distinct ways. 
First, the mRNA may be degraded, with its lifetime drawn from the exponential distribution specified in Eq.~\eqref{eq:mRNA_expo_distr}. Second, as mentioned above, translation can be interrupted when an elongating ribosome is stalled by antibiotic binding. Since the unbinding time of the antibiotic is typically much longer than the average mRNA lifetime, we assume that no additional proteins are produced after such a pausing event.
Consequently, for any given transcript, protein synthesis stops either when the mRNA degrades or a ribosome pauses.

To quantitatively describe the protein output per mRNA, we must first specify how protein numbers accumulate as a function of time. Given that the first protein is produced once the first ribosome crosses the lattice, the total number of proteins produced up to time $t$ is
\begin{equation}\label{eq:protein_output}
    \mathcal{P}(t) =
    \begin{cases}
        0, & t < \Tss \\
        J_0 (t - \Tss), & t \geq \Tss
    \end{cases}
\end{equation}
where $J_0$ is the steady-state translation current, see Eq.~(\ref{eq:current_standard}), and $\Tss$ denotes the onset of steady-state translation.
To determine the expected protein output per mRNA $\mathcal{P} = \int \mathcal{P}(t) \, \varphi_{\text{mRNA}}(t)\, dt$, we must consider both possible cases: mRNA degradation before pausing, or early ribosome pausing followed by transcript degradation before resumption of translation.

\textit{(i) Degradation before pausing.} 
For those mRNAs that, by chance, degrade before the first pausing event at time $\Tfp$, see Eq.~(\ref{eq:expected_time_until_pause}), translation proceeds until the degradation of the transcript occurs. The expected number of proteins produced by such mRNAs is given by
\begin{equation}
    \int_{\Tss}^{\Tfp} \mathcal{P}(t)\, \varphi_{\text{mRNA}}(t)\, dt,
\end{equation}
where this integral represents the amount of protein produced up to time $T_0$ (the onset of the paused state). It is obtained by weighting the protein output $\mathcal{P}(t)$ by the probability distribution $\varphi_{\text{mRNA}}(t)$, which describes the likelihood of mRNA degradation occurring at time $t$.

\textit{(ii) Pausing before degradation.} Conversely, consider the case in which an mRNAs has not yet degraded when the expected time of the first pausing event, $\Tfp$, is reached. 
In this case, translation is interrupted by antibiotic-induced pausing, and no further proteins are synthesized thereafter. All such mRNAs produce the same number of proteins, $\mathcal{P}(\Tfp)$, given by equation~\eqref{eq:protein_output}, evaluated at the first pausing time $\Tfp$. The total contribution from these mRNAs is
\begin{equation}
    \mathcal{P}(\Tfp) \int_{\Tfp}^{\infty} \varphi_{\text{mRNA}}(t)\, dt,  
\end{equation}
where the integral represents the weight of mRNAs that degrade after $\Tfp$. 

Combining both terms, the total expected number of proteins produced per mRNA, $\mathcal{P}$, is given by
\begin{align}\label{eq:N_proteins}
    \mathcal{P} &= \int_{\Tss}^{\Tfp} \mathcal{P}(t) \, \varphi_{\text{mRNA}}(t)\, dt \nonumber\\
    &\quad + \mathcal{P}(\Tfp) \int_{\Tfp}^{\infty} \, \varphi_{\text{mRNA}}(t)\, dt.
\end{align}
Evaluating Eq.~\eqref{eq:N_proteins} for the exponential mRNA lifetime distribution yields
\begin{equation}\label{eq:final_number_of_proteins}
    \mathcal{P} =
    \begin{cases}
        J_0\, \tau \left(e^{-\Tss/\tau} - e^{-\Tfp/\tau}\right), & \text{if } \Tfp > \Tss \\[2ex]
        0, & \text{if } \Tfp \leq \Tss
    \end{cases}
\end{equation}
This expression reflects the effective time window between $\Tss$ and $\Tfp$ during which protein synthesis occurs, weighted by the probability that the mRNA has not yet degraded. 
Note that the second case in Eq.~(\ref{eq:final_number_of_proteins}) is justified since  translation leads only to a negligible protein output if the pausing occurs before any ribosome completes.

\begin{figure}
    \centering
    \includegraphics[width=\columnwidth]{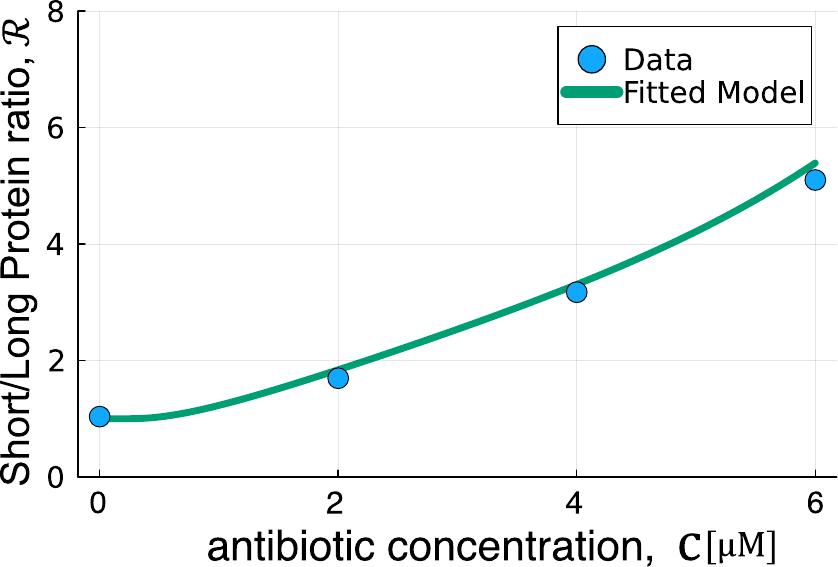}
    \caption{Comparison between model predictions and experimental data from Zhang et al.~\cite{zhang_decrease_2020} The parameters used are \(k_{\text{on}} = 5.4 \times 10^{-4}\,\mu M^{-1}s^{-1}\), \(k_u = 1.4\times10^{-3}\,s^{-1}\) (giving \(K_D \approx 2.5\,\mu M\)), with a fitted initiation rate of \(0.12\,s^{-1}\) and a mean mRNA lifetime \(\tau=170\,s\).}
    \label{fig:Experimental evidence}
\end{figure}

Because both reporters are driven by the same promoter at the same genomic locus, we assume any antibiotic‐induced change in mRNA abundance is identical for GFP and lacZ. By normalizing each protein’s expression under antibiotic treatment to its own untreated level, all effects of altered mRNA concentration cancel out—and only the relative change (fold-change) in protein output remains. Consequently, the ratio of fold-changes between the short (GFP) and long (lacZ) constructs,
\begin{equation}
    \mathcal{R} =
    \frac{ \mathcal{P}_s / \mathcal{P}_s^{(0)} }
         { \mathcal{P}_\ell / \mathcal{P}_\ell^{(0)} }
\end{equation}

reflects purely length-dependent translation inhibition, independent of transcript-level effects. Here, the indices \( s \) and \( \ell \) denote short and long proteins, respectively, and \((0)\) indicates the reference conditions without antibiotics.

The predictions from our model quantitatively match the experimental data, as shown in Fig.~\ref{fig:Experimental evidence}. Without fitting separate parameters for each gene (such as different initiation or elongation rates), the model captures the observed increase in the short-to-long expression ratio under increasing antibiotic concentrations. This agreement strongly supports the hypothesis that length-dependent inhibition naturally emerges from the interplay between collective ribosome dynamics and mRNA degradation, highlighting the model's predictive power. Furthermore, these results suggest that our framework can generalize to the expression of other genes under the same antibiotic, or to alternative inhibitory mechanisms affecting translation dynamics similarly to chloramphenicol.

\section{Discussion}

In this work, we developed a stochastic TASEP-based model to capture ribosome traffic under sub-lethal antibiotic stress, and that incorporates key biological constraints --finite mRNA length, transcript degradation, and experimentally measured antibiotic binding and unbinding rates. 
A strong timescale separation between translation dynamics and antibiotic binding/unbinding dynamics emerges from the experimental-derived parameters. This timescale separation enables us to develop a coarse-grained two-state description of translation consisting of an unpaused (translationally active mRNA) and a paused (translationally inhibited mRNA) state. This approach allows us to understand the interplay between ribosome-targeting antibiotic, protein synthesis, and mRNA features such as its lifetime, initiation rate, and its length.

%two-state
We thoroughly characterized these two states by focusing on the survival time of each and on the properties of translation (ribosome density and current) at different concentrations of antibiotics, i.e. as a function of the ribosome pausing rate $k_p$. A translationally active mRNA, free of ribosomes inactivated by antibiotics, transitions into an inhibited (paused) state upon antibiotic binding to a ribosome. The affected ribosome remains paused for an extended period, clogging the transcript and creating a downstream cluster of ribosomes.

%transition
As ribosomes continue to load onto the mRNA, the transition rate from the unpaused to the paused state increases linearly with time. At steady state, this transition rate is proportional to the number of ribosomes on the transcript and consequently to the mRNA length $L$. Therefore, longer transcripts and higher initiation rates accelerate ribosome accumulation, increasing the likelihood of early pausing events (Fig.~\ref{fig:time_until_first_pause}).

%the weights
After deriving the probability density function (pdf) for the survival time of the unpaused state $T_0$ and estimating the time $T_p$ required to resolve the initial cluster and revert to a state without paused ribosomes, we calculated the fraction of time the system spends in each state, as illustrated in Fig.~\ref{fig:weighted_and_unimpeded_translation}. 
The densities and currents are then calculated as weighted averages of the contributions from the unpaused (standard TASEP) and paused states.

%density and current
As the ribosome pausing rate $k_p$ increases, corresponding to higher antibiotic concentrations, Fig.~\ref{fig:density_and_current}(a) shows that the ribosome density initially rises sharply due to the formation of ribosome clusters and an increased probability of the system being in the paused state. 
Further increases in $k_p$ subsequently shift the initial paused ribosome closer to the initiation site (Fig.~\ref{fig:relative_pausing_position}), thereby reducing the total weighted average density.
On the contrary, the protein synthesis rate modeled by the current monotonically decreases with the antibiotic concentration, see Fig.~\ref{fig:density_and_current}(b).
Both observables severely depend on the transcript length.

%L-dependence is a collective effect
The dependence on system size arises as a collective effect, often not considered in previous studies. Such studies typically quantify the impact of translation-inhibiting antibiotics by estimating the probability of a single ribosome becoming stalled during mRNA translation. In other words, they focus on the likelihood of antibiotic binding during the translation time~$t_L$ of a \emph{single} ribosome~\cite{Dai2016, zhang_decrease_2020}. 
In contrast, our model incorporates the collective nature of ribosome traffic, revealing a mechanism where the probability of translation disruption increases with the number of ribosomes actively
translating the mRNA. As illustrated in Fig.~\ref{fig:hitting_probability}(a), this leads
to significantly higher hitting probabilities compared to predictions from single-particle models, especially for longer transcripts. Our model also predicts that longer mRNAs can reduce their vulnerability to ribosome-targeting antibiotics if they had lower initiation rates, thereby
decreasing ribosome density and exposure to pausing events, Fig.~\ref{fig:hitting_probability}(b). Interestingly, an anti-correlation between transcript lengths and initiation rates has been found in the literature, at least in yeast~\cite{fernandes_gene_2017}. 

%active particles
Moreover, as our theory explicitly addresses the impact of antibiotic binding and unbinding on overall ribosome dynamics and traffic, we introduce a formal definition of \emph{active} ribosomes, commonly inferred through phenomenological frameworks~\cite{Dai2016,Scott2010}. Specifically, our model demonstrates that defining active ribosomes strictly as those contributing to the current (mobile particles) yields results different from estimates commonly found in the literature, where ribosome interference (jammed particles) is usually neglected, Fig.~\ref{fig:particle_fractions}(a). However, we emphasize that, \emph{in vivo}, cells may utilize various regulatory mechanisms to prevent ribosome clustering~\cite{ikeuchi_recent_2019}.

%predictions and data comparisons
An important mechanism to prevent ribosome clustering and inefficient ribosome usage is mRNA degradation. The first pausing event significantly influences translational output under antibiotic stress since the mRNA lifetime is typically shorter than the chloramphenicol unbinding time. Thus, transcripts are often degraded before a paused ribosome can resume elongation, making protein production primarily occur before the first antibiotic-binding event.
This insight also explains the observed length-dependent effects of ribosome-targeting antibiotics on gene expression. Assuming an exponentially distributed mRNA lifetime with mean~$\tau$, our model quantitatively reproduces experimental differences in expression of \emph{lacZ} and \emph{gfp} (Fig.~\ref{fig:Experimental evidence}) by fitting only the degradation time~$\tau$, which falls within biologically plausible values. Consequently, shorter transcripts --such as those encoding many ribosomal proteins-- are predicted to be inherently more resistant to translation-inhibiting antibiotics compared to longer transcripts, such as those encoding RNA polymerase subunits.

%limitations
While our model captures key features of ribosome dynamics under antibiotic stress, it also has important limitations. The most important one is the assumption of a single paused cluster, which holds for slow-binding antibiotics like chloramphenicol but breaks down at higher binding rates, for instance, in the case of fusidic acid~\cite{uemura2010real}. Under such conditions, multiple paused clusters may form and interact through coagulation and decoagulation dynamics. Although the Single-Cluster Approximation loses predictive reliability once multiple paused clusters emerge, length-dependent effects are still expected to matter even in a refined approach to resolve this more complex behavior. 

%transcription
Despite simplifying assumptions, our framework naturally extends to other biological systems characterized by stochastic pausing~\cite{neuman_ubiquitous_2003, zhu_transcription_2022, janissen_high_throughput_2022}. A key example is transcription, where RNAPs intermittently pause, causing local traffic jams~\cite{Klumpp2008}. These intrinsic pauses typically occur at rates significantly slower than elongation. Notably, shorter genes often exhibit higher transcription initiation rates in bacteria  and yeast~\cite{balakrishnan_principles_2022, moreira_reporters_2019, shah_rate_limiting_2013}. Our conclusion that the density would depend on both gene length and initiation rate allows us to speculate that initiation rates have been adapted to minimise the onset of RNAP clustering, see Appendix~\ref{sec:app_transcription}).
We further emphasize that, as shown in~\cite{keisers2024finite}, our model of stochastically pausing particles can be mapped onto systems with dynamic obstacles~\cite{turci_transport_2013, waclaw_totally_2019} that temporarily impede particle movement. This analogy suggests broader applications, such as describing interactions between RNAPs and transcriptional regulatory elements or nucleosomes~\cite{van_crowding_2017, mines_slow_2022} or, beyond biology, systems such as vehicular traffic interrupted by traffic lights.
%future
Future work could extend this framework by connecting multiple mRNAs through a multi-TASEP approach~\cite{greulich_mixed_2012}, thereby linking antibiotic concentration to global bacterial outcomes such as growth rate and ribosome allocation.

\begin{acknowledgments}
J.K developed the theoretical framework and performed the simulations. L.C. conceptualized and supervised the work, supported by N.K. 
All authors discussed and interpreted the results and contributed to the final manuscript. We are grateful to Bianca Sclavi for sharing the experimental data used in Fig.~\ref{fig:Experimental evidence}. LC and JK were supported by the French National Research Agency (REF: ANR-21-CE45-0009).
\end{acknowledgments}

\appendix

\section{Simulations}
\label{app:sims}
We employ a standard Gillespie algorithm~\cite{gillespie1976general} for the simulations. The algorithm consists of two main phases. First, the system is allowed to run for a time $t_{ss}$ to reach steady state. After achieving steady state, data is collected for a duration $\Delta t$.

In the Gillespie algorithm, updates occur as Poisson processes, which means they are independent of each other. To ensure that the system reaches steady state and that the data collection period $\Delta t$ is sufficiently long, the total simulation time is made proportional to the lowest rate in the system. Unless otherwise stated,  we set $t_{ss} = 100 \times$ (lowest rate) and $\Delta t = 1000 \times$ (lowest rate). Thus, the total simulation time $t$ is equal $t=t_{ss} + \Delta t$.

\section{Finite‐cluster batch size calculation}
\label{sec:appendix_batch}

Here we derive the expression for the batch size \(B\) in a cluster of finite length. We begin by noting that the spacing between successive paused ribosomes can be viewed as the number of consecutive “free” (unpaused) particles preceding a paused one. Equivalently, this is a Bernoulli process in which a “success” (with probability \(p = 1 - f_a\)) corresponds to encountering a paused particle, and a “failure” (with probability \(f_a\)) to an unpaused particle.

Let \(x\) denote the number of consecutive unpaused particles upstream of a paused ribosome.  In the infinite‐cluster limit, \(x\) follows a geometric distribution with success probability \(1-f_a\), so that
\[
  \langle x\rangle
  = \sum_{i=0}^\infty (1-f_a)\,f_a^i\,i
  = \frac{f_a}{1 - f_a}
  = \frac{k_u}{k_p}.
\]

In a finite cluster of size \(N\), \(x\) is truncated at \(N-1\).  Splitting the expectation into two parts gives
\[
  d 
  = \sum_{i=0}^{N-2} (1-f_a)\,f_a^i\,i 
    + \sum_{i=N-1}^{\infty} (1-f_a)\,f_a^i\,(N-1).
\]
The first sum accounts for all allowed values \(0 \le i \le N-2\), while the second caps \(i\) at \(N-1\).  Evaluating these truncated geometric series yields
\[
  d 
  = \frac{f_a}{1 - f_a}\,\bigl[1 - f_a^N\bigr]
  = \frac{k_u}{k_p}\,\Bigl[1 - \bigl(\tfrac{k_p}{k_p + k_u}\bigr)^{N}\Bigr].
\]

Finally, since each batch also includes the paused ribosome at its head, the mean batch size is
\[\label{eq:supp_B}
  B = d + 1
    = \frac{k_u}{k_p}\Bigl[1 - \bigl(\tfrac{k_p}{k_p + k_u}\bigr)^{N}\Bigr] + 1.
\]
In the limit \(N\to\infty\), this recovers \(B = k_u/k_p + 1\).

\begin{figure}
    \centering
    \includegraphics[width=\columnwidth]{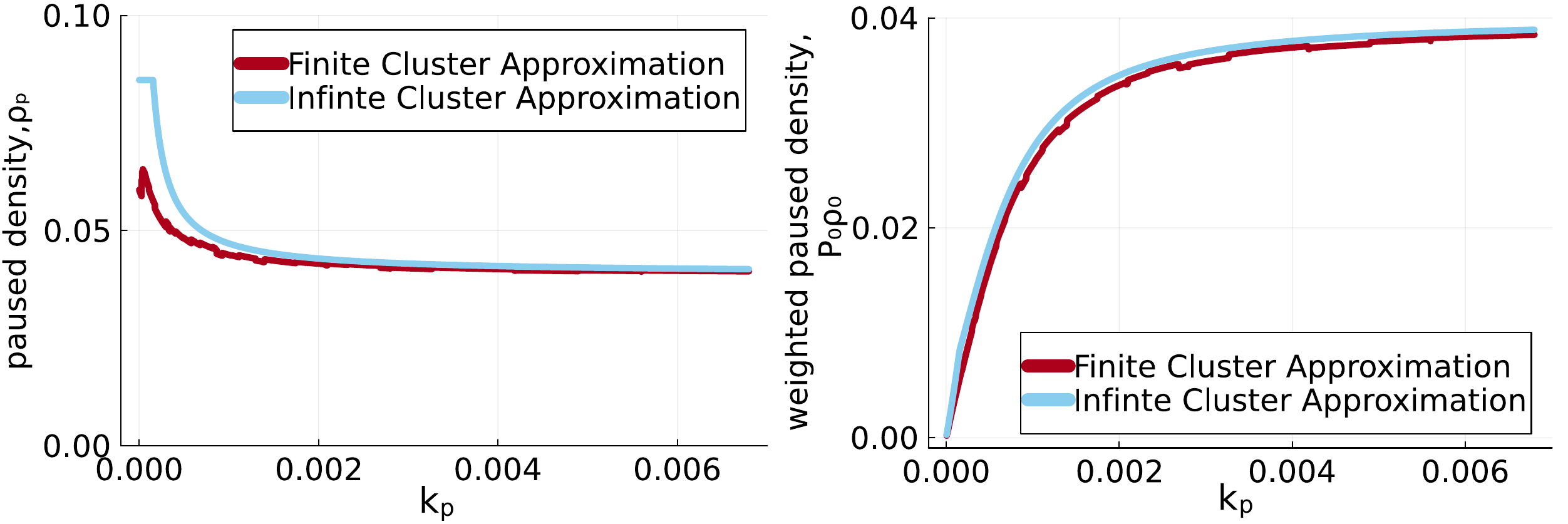}
    \caption{(a) Paused‐state density \(\rho_p\) for \(L=100\), comparing the finite‐cluster batch size \(B\) against the infinite‐cluster approximation \(B = k_u/k_p + 1\).  The finite‐cluster density \(\rho_p\) is obtained by numerically evaluating Eq.~\eqref{eq:density(N)} with the batch size depending on cluster size in mind.}
    \label{fig:app_density_in_paused_state}
\end{figure}

In order to obtain a closed-form expression for the density in Section~\ref{sec:particle_density}, we treat \(B\) as independent of the cluster size. However, if we account for the finite number of particles in the system --i.e., make \(B\) a function of the cluster size-- Eq.~\eqref{eq:density(N)} no longer admits a closed-form solution and must be solved numerically. Figure~\ref{fig:app_density_in_paused_state}(a) compares the unweighted paused-state density \(\rho_p\) computed using the finite-cluster expression for \(B\), via numerical solution of Eq.~\eqref{eq:density(N)}, with the infinite-cluster approximation \(B = k_u/k_p + 1\). As expected, the two agree closely for high pausing rates, while deviations become pronounced at low pausing rates. This discrepancy arises because, as \(k_p \to 0\), the batch size diverges. Note that, we limit the batch size to the maximum number of particles in the system $(L/ \ell)$. However, in the final density calculation, \(\rho_p\) is weighted by the probability of being in the paused state, which vanishes in this limit. These two effects approximately cancel out, making the infinite-cluster approximation a good one over the full range, see Fig.~\ref{fig:app_density_in_paused_state}(b).

\section{RNAP Pausing Dynamics and Gene Length-Dependent Initiation Rates}
\label{sec:app_transcription}

While our analysis has focused on translation, the TASEP framework with stochastic pausing is equally applicable to transcription. RNA polymerases (RNAPs) elongating along DNA also undergo random pauses, a phenomenon observed across species~\cite{neuman_ubiquitous_2003, janissen_high_throughput_2022}. These pauses arise from intrinsic mechanisms such as backtracking or proofreading~\cite{zhu_transcription_2022}, and typically occur on timescales two to three orders of magnitude slower than elongation~\cite{neuman_ubiquitous_2003, Klumpp2008}. Unlike pausing from external stress (e.g. antibiotics in translation), transcriptional pausing is an inherent feature of RNAP dynamics.

\begin{figure}[h] 
    \centering 
    \includegraphics[width=\linewidth]{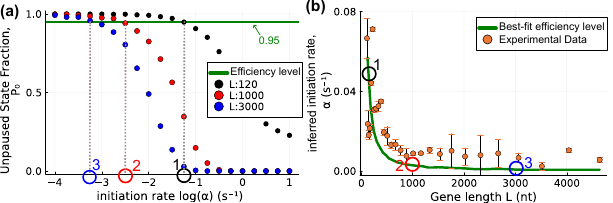} 
    \caption{(a) the Unpaused state fraction $P_0$ as a function of initiation rates, for different system sizes $L$. Here, we choose the efficiency level (green vertical line) so that the predicted initiation rate for a certain gene length matches the experimentally observed one, shown in (b). Experimental data is from~\cite{balakrishnan_principles_2022} (symbols) show a similar trend. Here, efficiency refers to the fraction of time spend in the unpaused state. The pausing and unpausing rate are taken from~\cite{Klumpp2008} and are here $k_u = 0.5$, $\varepsilon = 50$, $\ell = 30$ and $k_p = 0.15$.} 
    \label{fig:transcription} 
\end{figure}

Notably, experiments have found that shorter genes tend to exhibit higher transcription initiation rates in both bacteria and yeast~\cite{shah_rate_limiting_2013, balakrishnan_principles_2022, moreira_reporters_2019}. We hypothesize that this reflects an evolutionary adaptation to minimize RNAP congestion: longer genes, being more prone to queuing due to random pausing, require lower initiation rates to maintain efficient transcription. For example, at the same initiation rate, longer genes spend far more time in the unpaused state compared to short ones, as seen in figure~\ref{fig:transcription}(a).

To test this idea, we define transcriptional efficiency as the fraction of time the system spends in the unpaused state. For each gene length, we use simulations to determine the initiation rate that best matches the experimental data shown in Fig.~\ref{fig:transcription}(b). Our analysis reveals that reproducing the observed initiation rates requires transcription to remain in the unpaused state for over $90$ per cent of the time. This suggests that the initiation rates were optimize to reduce traffic and thus efficient RNAP usage, given their limited availability and high cellular cost. This analysis uncovers a clear trend: as gene length increases, the transcription process becomes more inefficient if the initiation rate does not change. This prediction aligns with experimental observations and supports the idea that stochastic pausing could imposes a physical constraint on transcription, shaping gene-specific initiation rates.

\bibliographystyle{apsrev4-2} % REVTeX recommended style
\bibliography{refined_citations}   

%apsrev4-2.bst 2019-01-14 (MD) hand-edited version of apsrev4-1.bst
%Control: key (0)
%Control: author (72) initials jnrlst
%Control: editor formatted (1) identically to author
%Control: production of article title (-1) disabled
%Control: page (0) single
%Control: year (1) truncated
%Control: production of eprint (0) enabled
\begin{thebibliography}{49}%
\makeatletter
\providecommand \@ifxundefined [1]{%
 \@ifx{#1\undefined}
}%
\providecommand \@ifnum [1]{%
 \ifnum #1\expandafter \@firstoftwo
 \else \expandafter \@secondoftwo
 \fi
}%
\providecommand \@ifx [1]{%
 \ifx #1\expandafter \@firstoftwo
 \else \expandafter \@secondoftwo
 \fi
}%
\providecommand \natexlab [1]{#1}%
\providecommand \enquote  [1]{``#1''}%
\providecommand \bibnamefont  [1]{#1}%
\providecommand \bibfnamefont [1]{#1}%
\providecommand \citenamefont [1]{#1}%
\providecommand \href@noop [0]{\@secondoftwo}%
\providecommand \href [0]{\begingroup \@sanitize@url \@href}%
\providecommand \@href[1]{\@@startlink{#1}\@@href}%
\providecommand \@@href[1]{\endgroup#1\@@endlink}%
\providecommand \@sanitize@url [0]{\catcode `\\12\catcode `\$12\catcode
  `\&12\catcode `\#12\catcode `\^12\catcode `\_12\catcode `\%12\relax}%
\providecommand \@@startlink[1]{}%
\providecommand \@@endlink[0]{}%
\providecommand \url  [0]{\begingroup\@sanitize@url \@url }%
\providecommand \@url [1]{\endgroup\@href {#1}{\urlprefix }}%
\providecommand \urlprefix  [0]{URL }%
\providecommand \Eprint [0]{\href }%
\providecommand \doibase [0]{https://doi.org/}%
\providecommand \selectlanguage [0]{\@gobble}%
\providecommand \bibinfo  [0]{\@secondoftwo}%
\providecommand \bibfield  [0]{\@secondoftwo}%
\providecommand \translation [1]{[#1]}%
\providecommand \BibitemOpen [0]{}%
\providecommand \bibitemStop [0]{}%
\providecommand \bibitemNoStop [0]{.\EOS\space}%
\providecommand \EOS [0]{\spacefactor3000\relax}%
\providecommand \BibitemShut  [1]{\csname bibitem#1\endcsname}%
\let\auto@bib@innerbib\@empty
%</preamble>
\bibitem [{\citenamefont {Klumpp}\ and\ \citenamefont {Hwa}()}]{Klumpp2008}%
  \BibitemOpen
  \bibfield  {author} {\bibinfo {author} {\bibfnamefont {S.}~\bibnamefont
  {Klumpp}}\ and\ \bibinfo {author} {\bibfnamefont {T.}~\bibnamefont {Hwa}}\
  }\textbf {\bibinfo {volume} {105}},\ \href
  {https://doi.org/10.1073/pnas.0806084105}
  {10.1073/pnas.0806084105}\BibitemShut {NoStop}%
\bibitem [{\citenamefont {Kavčič}\ \emph {et~al.}()\citenamefont {Kavčič},
  \citenamefont {Tkačik},\ and\ \citenamefont {Bollenbach}}]{Kavcic2020}%
  \BibitemOpen
  \bibfield  {author} {\bibinfo {author} {\bibfnamefont {B.}~\bibnamefont
  {Kavčič}}, \bibinfo {author} {\bibfnamefont {G.}~\bibnamefont {Tkačik}},\
  and\ \bibinfo {author} {\bibfnamefont {T.}~\bibnamefont {Bollenbach}}\
  }\textbf {\bibinfo {volume} {11}},\ \href
  {https://doi.org/10.1038/s41467-020-17734-z}
  {10.1038/s41467-020-17734-z}\BibitemShut {NoStop}%
\bibitem [{\citenamefont {MacDonald}\ \emph {et~al.}(1968)\citenamefont
  {MacDonald}, \citenamefont {Gibbs},\ and\ \citenamefont
  {Pipkin}}]{MacDonald1968}%
  \BibitemOpen
  \bibfield  {author} {\bibinfo {author} {\bibfnamefont {C.~T.}\ \bibnamefont
  {MacDonald}}, \bibinfo {author} {\bibfnamefont {J.~H.}\ \bibnamefont
  {Gibbs}},\ and\ \bibinfo {author} {\bibfnamefont {A.~C.}\ \bibnamefont
  {Pipkin}},\ }\href {https://doi.org/10.1002/bip.1968.360060102} {\bibfield
  {journal} {\bibinfo  {journal} {Biopolymers}\ }\textbf {\bibinfo {volume}
  {6}},\ \bibinfo {pages} {1} (\bibinfo {year} {1968})}\BibitemShut {NoStop}%
\bibitem [{\citenamefont {Blythe}\ and\ \citenamefont
  {Evans}(2007)}]{Blythe2007}%
  \BibitemOpen
  \bibfield  {author} {\bibinfo {author} {\bibfnamefont {R.~A.}\ \bibnamefont
  {Blythe}}\ and\ \bibinfo {author} {\bibfnamefont {M.~R.}\ \bibnamefont
  {Evans}},\ }\bibfield  {journal} {\bibinfo  {journal} {Journal of Physics A:
  Mathematical and Theoretical}\ }\textbf {\bibinfo {volume} {40}},\ \href
  {https://doi.org/10.1088/1751-8113/40/46/R01} {10.1088/1751-8113/40/46/R01}
  (\bibinfo {year} {2007})\BibitemShut {NoStop}%
\bibitem [{\citenamefont {Tripathi}\ and\ \citenamefont
  {Chowdhury}(2008)}]{tripathi2008interacting}%
  \BibitemOpen
  \bibfield  {author} {\bibinfo {author} {\bibfnamefont {T.}~\bibnamefont
  {Tripathi}}\ and\ \bibinfo {author} {\bibfnamefont {D.}~\bibnamefont
  {Chowdhury}},\ }\href {https://doi.org/10.1103/PhysRevE.77.011921} {\bibfield
   {journal} {\bibinfo  {journal} {Phys. Rev. E}\ }\textbf {\bibinfo {volume}
  {77}},\ \bibinfo {pages} {011921} (\bibinfo {year} {2008})}\BibitemShut
  {NoStop}%
\bibitem [{\citenamefont {Wang}\ \emph {et~al.}()\citenamefont {Wang},
  \citenamefont {Pfeuty}, \citenamefont {Thommen}, \citenamefont {Romano},\
  and\ \citenamefont {Lefranc}}]{Wang_minimal_2014}%
  \BibitemOpen
  \bibfield  {author} {\bibinfo {author} {\bibfnamefont {J.}~\bibnamefont
  {Wang}}, \bibinfo {author} {\bibfnamefont {B.}~\bibnamefont {Pfeuty}},
  \bibinfo {author} {\bibfnamefont {Q.}~\bibnamefont {Thommen}}, \bibinfo
  {author} {\bibfnamefont {M.~C.}\ \bibnamefont {Romano}},\ and\ \bibinfo
  {author} {\bibfnamefont {M.}~\bibnamefont {Lefranc}}\ }\textbf {\bibinfo
  {volume} {90}},\ \href {https://doi.org/10.1103/PhysRevE.90.050701}
  {10.1103/PhysRevE.90.050701}\BibitemShut {NoStop}%
\bibitem [{\citenamefont {van~den Berg}\ and\ \citenamefont
  {Depken}(2017)}]{van_crowding_2017}%
  \BibitemOpen
  \bibfield  {author} {\bibinfo {author} {\bibfnamefont {A.~A.}\ \bibnamefont
  {van~den Berg}}\ and\ \bibinfo {author} {\bibfnamefont {M.}~\bibnamefont
  {Depken}},\ }\href@noop {} {\bibfield  {journal} {\bibinfo  {journal}
  {Nucleic Acids Research}\ }\textbf {\bibinfo {volume} {45}},\ \bibinfo
  {pages} {7623} (\bibinfo {year} {2017})}\BibitemShut {NoStop}%
\bibitem [{\citenamefont {Mines}\ \emph {et~al.}(2022)\citenamefont {Mines},
  \citenamefont {Lipniacki},\ and\ \citenamefont {Shen}}]{mines_slow_2022}%
  \BibitemOpen
  \bibfield  {author} {\bibinfo {author} {\bibfnamefont {R.~C.}\ \bibnamefont
  {Mines}}, \bibinfo {author} {\bibfnamefont {T.}~\bibnamefont {Lipniacki}},\
  and\ \bibinfo {author} {\bibfnamefont {X.}~\bibnamefont {Shen}},\ }\href
  {https://doi.org/10.1371/journal.pcbi.1009811} {\bibfield  {journal}
  {\bibinfo  {journal} {PLOS Computational Biology}\ }\textbf {\bibinfo
  {volume} {18}},\ \bibinfo {pages} {1} (\bibinfo {year} {2022})}\BibitemShut
  {NoStop}%
\bibitem [{\citenamefont {Shaw}\ \emph {et~al.}(2003)\citenamefont {Shaw},
  \citenamefont {Zia},\ and\ \citenamefont {Lee}}]{shaw_totally_2003}%
  \BibitemOpen
  \bibfield  {author} {\bibinfo {author} {\bibfnamefont {L.~B.}\ \bibnamefont
  {Shaw}}, \bibinfo {author} {\bibfnamefont {R.~K.~P.}\ \bibnamefont {Zia}},\
  and\ \bibinfo {author} {\bibfnamefont {K.~H.}\ \bibnamefont {Lee}},\ }\href
  {https://doi.org/10.1103/PhysRevE.68.021910} {\bibfield  {journal} {\bibinfo
  {journal} {Physical Review E}\ }\textbf {\bibinfo {volume} {68}},\ \bibinfo
  {pages} {021910} (\bibinfo {year} {2003})}\BibitemShut {NoStop}%
\bibitem [{\citenamefont {Klumpp}\ \emph {et~al.}(2008)\citenamefont {Klumpp},
  \citenamefont {Chai},\ and\ \citenamefont {Lipowsky}}]{klumpp_effects_2008}%
  \BibitemOpen
  \bibfield  {author} {\bibinfo {author} {\bibfnamefont {S.}~\bibnamefont
  {Klumpp}}, \bibinfo {author} {\bibfnamefont {Y.}~\bibnamefont {Chai}},\ and\
  \bibinfo {author} {\bibfnamefont {R.}~\bibnamefont {Lipowsky}},\ }\href
  {https://doi.org/10.1103/PhysRevE.78.041909} {\bibfield  {journal} {\bibinfo
  {journal} {Physical Review E}\ }\textbf {\bibinfo {volume} {78}},\ \bibinfo
  {pages} {041909} (\bibinfo {year} {2008})}\BibitemShut {NoStop}%
\bibitem [{\citenamefont {Ciandrini}\ \emph {et~al.}(2010)\citenamefont
  {Ciandrini}, \citenamefont {Stansfield},\ and\ \citenamefont
  {Romano}}]{ciandrini_role_2010}%
  \BibitemOpen
  \bibfield  {author} {\bibinfo {author} {\bibfnamefont {L.}~\bibnamefont
  {Ciandrini}}, \bibinfo {author} {\bibfnamefont {I.}~\bibnamefont
  {Stansfield}},\ and\ \bibinfo {author} {\bibfnamefont {M.~C.}\ \bibnamefont
  {Romano}},\ }\href {https://doi.org/10.1103/PhysRevE.81.051904} {\bibfield
  {journal} {\bibinfo  {journal} {Physical Review E}\ }\textbf {\bibinfo
  {volume} {81}},\ \bibinfo {pages} {051904} (\bibinfo {year}
  {2010})}\BibitemShut {NoStop}%
\bibitem [{\citenamefont {Rousset}\ \emph {et~al.}(2019)\citenamefont
  {Rousset}, \citenamefont {Ciandrini},\ and\ \citenamefont
  {Kern}}]{Rousset2019}%
  \BibitemOpen
  \bibfield  {author} {\bibinfo {author} {\bibfnamefont {Y.}~\bibnamefont
  {Rousset}}, \bibinfo {author} {\bibfnamefont {L.}~\bibnamefont {Ciandrini}},\
  and\ \bibinfo {author} {\bibfnamefont {N.}~\bibnamefont {Kern}},\ }\href
  {https://doi.org/10.21468/SciPostPhys.6.6.077} {\bibfield  {journal}
  {\bibinfo  {journal} {SciPost Phys.}\ }\textbf {\bibinfo {volume} {6}},\
  \bibinfo {pages} {077} (\bibinfo {year} {2019})}\BibitemShut {NoStop}%
\bibitem [{\citenamefont {Shaw}\ \emph {et~al.}(2004)\citenamefont {Shaw},
  \citenamefont {Sethna},\ and\ \citenamefont {Lee}}]{shaw_mean_2004}%
  \BibitemOpen
  \bibfield  {author} {\bibinfo {author} {\bibfnamefont {L.~B.}\ \bibnamefont
  {Shaw}}, \bibinfo {author} {\bibfnamefont {J.~P.}\ \bibnamefont {Sethna}},\
  and\ \bibinfo {author} {\bibfnamefont {K.~H.}\ \bibnamefont {Lee}},\
  }\href@noop {} {\bibfield  {journal} {\bibinfo  {journal} {Physical Review
  E—Statistical, Nonlinear, and Soft Matter Physics}\ }\textbf {\bibinfo
  {volume} {70}},\ \bibinfo {pages} {021901} (\bibinfo {year}
  {2004})}\BibitemShut {NoStop}%
\bibitem [{\citenamefont {Szavits-Nossan}\ \emph {et~al.}(2018)\citenamefont
  {Szavits-Nossan}, \citenamefont {Ciandrini},\ and\ \citenamefont
  {Romano}}]{szavits2018deciphering}%
  \BibitemOpen
  \bibfield  {author} {\bibinfo {author} {\bibfnamefont {J.}~\bibnamefont
  {Szavits-Nossan}}, \bibinfo {author} {\bibfnamefont {L.}~\bibnamefont
  {Ciandrini}},\ and\ \bibinfo {author} {\bibfnamefont {M.~C.}\ \bibnamefont
  {Romano}},\ }\href@noop {} {\bibfield  {journal} {\bibinfo  {journal} {Phys.
  Rev. Lett.}\ }\textbf {\bibinfo {volume} {120}},\ \bibinfo {pages} {128101}
  (\bibinfo {year} {2018})}\BibitemShut {NoStop}%
\bibitem [{\citenamefont {Erdmann-Pham}\ \emph {et~al.}(2020)\citenamefont
  {Erdmann-Pham}, \citenamefont {Duc},\ and\ \citenamefont
  {Song}}]{erdmann2020key}%
  \BibitemOpen
  \bibfield  {author} {\bibinfo {author} {\bibfnamefont {D.~D.}\ \bibnamefont
  {Erdmann-Pham}}, \bibinfo {author} {\bibfnamefont {K.~D.}\ \bibnamefont
  {Duc}},\ and\ \bibinfo {author} {\bibfnamefont {Y.~S.}\ \bibnamefont
  {Song}},\ }\href@noop {} {\bibfield  {journal} {\bibinfo  {journal} {Cell
  systems}\ }\textbf {\bibinfo {volume} {10}},\ \bibinfo {pages} {183}
  (\bibinfo {year} {2020})}\BibitemShut {NoStop}%
\bibitem [{\citenamefont {Ciandrini}\ \emph {et~al.}(2023)\citenamefont
  {Ciandrini}, \citenamefont {Crisostomo},\ and\ \citenamefont
  {Szavits-Nossan}}]{ciandrini_tasepy_2023}%
  \BibitemOpen
  \bibfield  {author} {\bibinfo {author} {\bibfnamefont {L.}~\bibnamefont
  {Ciandrini}}, \bibinfo {author} {\bibfnamefont {R.~L.}\ \bibnamefont
  {Crisostomo}},\ and\ \bibinfo {author} {\bibfnamefont {J.}~\bibnamefont
  {Szavits-Nossan}},\ }\href {https://doi.org/10.21468/SciPostPhysCodeb.22}
  {\bibfield  {journal} {\bibinfo  {journal} {SciPost Physics Codebases}\ ,\
  \bibinfo {pages} {022}} (\bibinfo {year} {2023})}\BibitemShut {NoStop}%
\bibitem [{\citenamefont {Turci}\ \emph {et~al.}(2013)\citenamefont {Turci},
  \citenamefont {Parmeggiani}, \citenamefont {Pitard}, \citenamefont {Romano},\
  and\ \citenamefont {Ciandrini}}]{turci_transport_2013}%
  \BibitemOpen
  \bibfield  {author} {\bibinfo {author} {\bibfnamefont {F.}~\bibnamefont
  {Turci}}, \bibinfo {author} {\bibfnamefont {A.}~\bibnamefont {Parmeggiani}},
  \bibinfo {author} {\bibfnamefont {E.}~\bibnamefont {Pitard}}, \bibinfo
  {author} {\bibfnamefont {M.~C.}\ \bibnamefont {Romano}},\ and\ \bibinfo
  {author} {\bibfnamefont {L.}~\bibnamefont {Ciandrini}},\ }\href
  {https://doi.org/10.1103/PhysRevE.87.012705} {\bibfield  {journal} {\bibinfo
  {journal} {Physical Review E}\ }\textbf {\bibinfo {volume} {87}},\ \bibinfo
  {pages} {012705} (\bibinfo {year} {2013})}\BibitemShut {NoStop}%
\bibitem [{\citenamefont {Waclaw}\ \emph {et~al.}()\citenamefont {Waclaw},
  \citenamefont {Cholewa-Waclaw},\ and\ \citenamefont
  {Greulich}}]{waclaw_totally_2019}%
  \BibitemOpen
  \bibfield  {author} {\bibinfo {author} {\bibfnamefont {B.}~\bibnamefont
  {Waclaw}}, \bibinfo {author} {\bibfnamefont {J.}~\bibnamefont
  {Cholewa-Waclaw}},\ and\ \bibinfo {author} {\bibfnamefont {P.}~\bibnamefont
  {Greulich}},\ }\href@noop {} {\ \textbf {\bibinfo {volume} {52}}}\BibitemShut
  {NoStop}%
\bibitem [{\citenamefont {Zhu}\ \emph {et~al.}(2022)\citenamefont {Zhu},
  \citenamefont {Guo}, \citenamefont {Dumas}, \citenamefont {Takacs},
  \citenamefont {Abdelkareem}, \citenamefont {Vanden~Broeck}, \citenamefont
  {Saint-André}, \citenamefont {Papai}, \citenamefont {Crucifix},
  \citenamefont {Ortiz},\ and\ \citenamefont
  {Weixlbaumer}}]{zhu_transcription_2022}%
  \BibitemOpen
  \bibfield  {author} {\bibinfo {author} {\bibfnamefont {C.}~\bibnamefont
  {Zhu}}, \bibinfo {author} {\bibfnamefont {X.}~\bibnamefont {Guo}}, \bibinfo
  {author} {\bibfnamefont {P.}~\bibnamefont {Dumas}}, \bibinfo {author}
  {\bibfnamefont {M.}~\bibnamefont {Takacs}}, \bibinfo {author} {\bibfnamefont
  {M.}~\bibnamefont {Abdelkareem}}, \bibinfo {author} {\bibfnamefont
  {A.}~\bibnamefont {Vanden~Broeck}}, \bibinfo {author} {\bibfnamefont
  {C.}~\bibnamefont {Saint-André}}, \bibinfo {author} {\bibfnamefont
  {G.}~\bibnamefont {Papai}}, \bibinfo {author} {\bibfnamefont
  {C.}~\bibnamefont {Crucifix}}, \bibinfo {author} {\bibfnamefont
  {J.}~\bibnamefont {Ortiz}},\ and\ \bibinfo {author} {\bibfnamefont
  {A.}~\bibnamefont {Weixlbaumer}},\ }\href
  {https://doi.org/10.1038/s41467-022-29148-0} {\bibfield  {journal} {\bibinfo
  {journal} {Nature Communications}\ }\textbf {\bibinfo {volume} {13}},\
  \bibinfo {pages} {1546} (\bibinfo {year} {2022})}\BibitemShut {NoStop}%
\bibitem [{\citenamefont {Zhang}\ \emph {et~al.}(2020)\citenamefont {Zhang},
  \citenamefont {Brambilla}, \citenamefont {Li}, \citenamefont {Shi},
  \citenamefont {Cosentino~Lagomarsino},\ and\ \citenamefont
  {Sclavi}}]{zhang_decrease_2020}%
  \BibitemOpen
  \bibfield  {author} {\bibinfo {author} {\bibfnamefont {Q.}~\bibnamefont
  {Zhang}}, \bibinfo {author} {\bibfnamefont {E.}~\bibnamefont {Brambilla}},
  \bibinfo {author} {\bibfnamefont {R.}~\bibnamefont {Li}}, \bibinfo {author}
  {\bibfnamefont {H.}~\bibnamefont {Shi}}, \bibinfo {author} {\bibfnamefont
  {M.}~\bibnamefont {Cosentino~Lagomarsino}},\ and\ \bibinfo {author}
  {\bibfnamefont {B.}~\bibnamefont {Sclavi}},\ }\bibfield  {journal} {\bibinfo
  {journal} {mSystems}\ }\textbf {\bibinfo {volume} {5}},\ \href
  {https://doi.org/10.1128/mSystems.00575-20} {10.1128/mSystems.00575-20}
  (\bibinfo {year} {2020})\BibitemShut {NoStop}%
\bibitem [{\citenamefont {Walsh}(2004)}]{Walsh2004}%
  \BibitemOpen
  \bibfield  {author} {\bibinfo {author} {\bibfnamefont {C.~T.}\ \bibnamefont
  {Walsh}},\ }\href {https://doi.org/10.1110/ps.041032204} {\bibfield
  {journal} {\bibinfo  {journal} {Protein Science}\ }\textbf {\bibinfo {volume}
  {13}},\ \bibinfo {pages} {3059} (\bibinfo {year} {2004})}\BibitemShut
  {NoStop}%
\bibitem [{\citenamefont {Harvey}\ and\ \citenamefont
  {Koch}()}]{harvey_how_1980}%
  \BibitemOpen
  \bibfield  {author} {\bibinfo {author} {\bibfnamefont {R.~J.}\ \bibnamefont
  {Harvey}}\ and\ \bibinfo {author} {\bibfnamefont {A.~L.}\ \bibnamefont
  {Koch}}\ }\textbf {\bibinfo {volume} {18}},\ \href
  {https://doi.org/10.1016/0022-2836(74)90234-4}
  {10.1016/0022-2836(74)90234-4}\BibitemShut {NoStop}%
\bibitem [{\citenamefont {Chopra}\ and\ \citenamefont
  {Howe}(1978)}]{Chopra_1978}%
  \BibitemOpen
  \bibfield  {author} {\bibinfo {author} {\bibfnamefont {I.}~\bibnamefont
  {Chopra}}\ and\ \bibinfo {author} {\bibfnamefont {T.~G.}\ \bibnamefont
  {Howe}},\ }\href@noop {} {\bibfield  {journal} {\bibinfo  {journal}
  {Microbiological Reviews}\ }\textbf {\bibinfo {volume} {42}},\ \bibinfo
  {pages} {707} (\bibinfo {year} {1978})}\BibitemShut {NoStop}%
\bibitem [{\citenamefont {Berens}\ \emph {et~al.}(2001)\citenamefont {Berens},
  \citenamefont {Thain},\ and\ \citenamefont
  {Schroeder}}]{berens_tetracycline-binding_2001}%
  \BibitemOpen
  \bibfield  {author} {\bibinfo {author} {\bibfnamefont {C.}~\bibnamefont
  {Berens}}, \bibinfo {author} {\bibfnamefont {A.}~\bibnamefont {Thain}},\ and\
  \bibinfo {author} {\bibfnamefont {R.}~\bibnamefont {Schroeder}},\ }\href
  {https://doi.org/10.1016/S0968-0896(01)00063-3} {\bibfield  {journal}
  {\bibinfo  {journal} {Bioorganic \& Medicinal Chemistry}\ }\textbf {\bibinfo
  {volume} {9}},\ \bibinfo {pages} {2549} (\bibinfo {year} {2001})}\BibitemShut
  {NoStop}%
\bibitem [{\citenamefont {Pestka}(1974)}]{Pestka1974}%
  \BibitemOpen
  \bibfield  {author} {\bibinfo {author} {\bibfnamefont {S.}~\bibnamefont
  {Pestka}},\ }\href@noop {} {\bibfield  {journal} {\bibinfo  {journal}
  {Antimicrobial Agents and Chemotherapy}\ }\textbf {\bibinfo {volume} {6}},\
  \bibinfo {pages} {474} (\bibinfo {year} {1974})}\BibitemShut {NoStop}%
\bibitem [{\citenamefont {Dai}\ \emph {et~al.}()\citenamefont {Dai},
  \citenamefont {Zhu}, \citenamefont {Warren}, \citenamefont {Balakrishnan},
  \citenamefont {Patsalo}, \citenamefont {Okano}, \citenamefont {Williamson},
  \citenamefont {Fredrick}, \citenamefont {Wang},\ and\ \citenamefont
  {Hwa}}]{Dai2016}%
  \BibitemOpen
  \bibfield  {author} {\bibinfo {author} {\bibfnamefont {X.}~\bibnamefont
  {Dai}}, \bibinfo {author} {\bibfnamefont {M.}~\bibnamefont {Zhu}}, \bibinfo
  {author} {\bibfnamefont {M.}~\bibnamefont {Warren}}, \bibinfo {author}
  {\bibfnamefont {R.}~\bibnamefont {Balakrishnan}}, \bibinfo {author}
  {\bibfnamefont {V.}~\bibnamefont {Patsalo}}, \bibinfo {author} {\bibfnamefont
  {H.}~\bibnamefont {Okano}}, \bibinfo {author} {\bibfnamefont {J.~R.}\
  \bibnamefont {Williamson}}, \bibinfo {author} {\bibfnamefont
  {K.}~\bibnamefont {Fredrick}}, \bibinfo {author} {\bibfnamefont {Y.-P.}\
  \bibnamefont {Wang}},\ and\ \bibinfo {author} {\bibfnamefont
  {T.}~\bibnamefont {Hwa}}\ }\textbf {\bibinfo {volume} {2}},\ \href
  {https://doi.org/10.1038/nmicrobiol.2016.231}
  {10.1038/nmicrobiol.2016.231}\BibitemShut {NoStop}%
\bibitem [{\citenamefont {Ciandrini}\ \emph {et~al.}(2013)\citenamefont
  {Ciandrini}, \citenamefont {Stansfield},\ and\ \citenamefont
  {Romano}}]{Ciandrini2013}%
  \BibitemOpen
  \bibfield  {author} {\bibinfo {author} {\bibfnamefont {L.}~\bibnamefont
  {Ciandrini}}, \bibinfo {author} {\bibfnamefont {I.}~\bibnamefont
  {Stansfield}},\ and\ \bibinfo {author} {\bibfnamefont {M.~C.}\ \bibnamefont
  {Romano}},\ }\href {https://doi.org/10.1371/journal.pcbi.1002866} {\bibfield
  {journal} {\bibinfo  {journal} {PLoS Comput. Biol.}\ }\textbf {\bibinfo
  {volume} {9}},\ \bibinfo {pages} {e1002866} (\bibinfo {year}
  {2013})}\BibitemShut {NoStop}%
\bibitem [{\citenamefont {Milo}\ and\ \citenamefont
  {Phillips}(2015)}]{milo2015cell}%
  \BibitemOpen
  \bibfield  {author} {\bibinfo {author} {\bibfnamefont {R.}~\bibnamefont
  {Milo}}\ and\ \bibinfo {author} {\bibfnamefont {R.}~\bibnamefont
  {Phillips}},\ }\href@noop {} {\emph {\bibinfo {title} {Cell Biology by the
  Numbers}}}\ (\bibinfo  {publisher} {Garland Science},\ \bibinfo {address}
  {New York},\ \bibinfo {year} {2015})\BibitemShut {NoStop}%
\bibitem [{\citenamefont {Balakrishnan}\ \emph {et~al.}(2022)\citenamefont
  {Balakrishnan}, \citenamefont {Mori}, \citenamefont {Segota}, \citenamefont
  {Zhang}, \citenamefont {Aebersold}, \citenamefont {Ludwig},\ and\
  \citenamefont {Hwa}}]{balakrishnan_principles_2022}%
  \BibitemOpen
  \bibfield  {author} {\bibinfo {author} {\bibfnamefont {R.}~\bibnamefont
  {Balakrishnan}}, \bibinfo {author} {\bibfnamefont {M.}~\bibnamefont {Mori}},
  \bibinfo {author} {\bibfnamefont {I.}~\bibnamefont {Segota}}, \bibinfo
  {author} {\bibfnamefont {Z.}~\bibnamefont {Zhang}}, \bibinfo {author}
  {\bibfnamefont {R.}~\bibnamefont {Aebersold}}, \bibinfo {author}
  {\bibfnamefont {C.}~\bibnamefont {Ludwig}},\ and\ \bibinfo {author}
  {\bibfnamefont {T.}~\bibnamefont {Hwa}},\ }\href
  {https://doi.org/10.1126/science.abk2066} {\bibfield  {journal} {\bibinfo
  {journal} {Science}\ }\textbf {\bibinfo {volume} {378}},\ \bibinfo {pages}
  {eabk2066} (\bibinfo {year} {2022})}\BibitemShut {NoStop}%
\bibitem [{\citenamefont {Keisers}\ \emph {et~al.}(2024)\citenamefont
  {Keisers}, \citenamefont {Dal~Zovo}, \citenamefont {Kern},\ and\
  \citenamefont {Ciandrini}}]{keisers2024finite}%
  \BibitemOpen
  \bibfield  {author} {\bibinfo {author} {\bibfnamefont {J.}~\bibnamefont
  {Keisers}}, \bibinfo {author} {\bibfnamefont {L.~V.}\ \bibnamefont
  {Dal~Zovo}}, \bibinfo {author} {\bibfnamefont {N.}~\bibnamefont {Kern}},\
  and\ \bibinfo {author} {\bibfnamefont {L.}~\bibnamefont {Ciandrini}},\ }\href
  {https://doi.org/10.48550/arXiv.2406.16569} {\bibinfo {title} {Biologically
  relevant finite-size effects in a driven lattice gas with particle pausing
  and dynamical defects}} (\bibinfo {year} {2024}),\ \Eprint
  {https://arxiv.org/abs/2406.16569} {arXiv:2406.16569 [cond-mat.stat-mech]}
  \BibitemShut {NoStop}%
\bibitem [{\citenamefont {Kolomeisky}\ \emph {et~al.}(1998)\citenamefont
  {Kolomeisky}, \citenamefont {Schütz}, \citenamefont {Kolomeisky},\ and\
  \citenamefont {Straley}}]{kolomeisky_phase_1998}%
  \BibitemOpen
  \bibfield  {author} {\bibinfo {author} {\bibfnamefont {A.~B.}\ \bibnamefont
  {Kolomeisky}}, \bibinfo {author} {\bibfnamefont {G.~M.}\ \bibnamefont
  {Schütz}}, \bibinfo {author} {\bibfnamefont {E.~B.}\ \bibnamefont
  {Kolomeisky}},\ and\ \bibinfo {author} {\bibfnamefont {J.~P.}\ \bibnamefont
  {Straley}},\ }\href {https://doi.org/10.1088/0305-4470/31/33/003} {\bibfield
  {journal} {\bibinfo  {journal} {Journal of Physics A: Mathematical and
  General}\ }\textbf {\bibinfo {volume} {31}},\ \bibinfo {pages} {6911}
  (\bibinfo {year} {1998})}\BibitemShut {NoStop}%
\bibitem [{\citenamefont {Santen}\ and\ \citenamefont
  {Appert}(2002)}]{santen_asymmetric_2002}%
  \BibitemOpen
  \bibfield  {author} {\bibinfo {author} {\bibfnamefont {L.}~\bibnamefont
  {Santen}}\ and\ \bibinfo {author} {\bibfnamefont {C.}~\bibnamefont
  {Appert}},\ }\href@noop {} {\bibfield  {journal} {\bibinfo  {journal}
  {Journal of statistical physics}\ }\textbf {\bibinfo {volume} {106}},\
  \bibinfo {pages} {187} (\bibinfo {year} {2002})}\BibitemShut {NoStop}%
\bibitem [{\citenamefont {Cividini}\ \emph {et~al.}(2014)\citenamefont
  {Cividini}, \citenamefont {Hilhorst},\ and\ \citenamefont
  {Appert-Rolland}}]{cividini_exact_2014}%
  \BibitemOpen
  \bibfield  {author} {\bibinfo {author} {\bibfnamefont {J.}~\bibnamefont
  {Cividini}}, \bibinfo {author} {\bibfnamefont {H.~J.}\ \bibnamefont
  {Hilhorst}},\ and\ \bibinfo {author} {\bibfnamefont {C.}~\bibnamefont
  {Appert-Rolland}},\ }\href {https://doi.org/10.1088/1751-8113/47/22/222001}
  {\bibfield  {journal} {\bibinfo  {journal} {Journal of Physics A:
  Mathematical and Theoretical}\ }\textbf {\bibinfo {volume} {47}},\ \bibinfo
  {pages} {222001} (\bibinfo {year} {2014})}\BibitemShut {NoStop}%
\bibitem [{\citenamefont {Szavits-Nossan}\ and\ \citenamefont
  {Evans}(2020)}]{szavits-nossan_dynamics_2020}%
  \BibitemOpen
  \bibfield  {author} {\bibinfo {author} {\bibfnamefont {J.}~\bibnamefont
  {Szavits-Nossan}}\ and\ \bibinfo {author} {\bibfnamefont {M.~R.}\
  \bibnamefont {Evans}},\ }\bibfield  {journal} {\bibinfo  {journal} {Physical
  Review E}\ }\textbf {\bibinfo {volume} {101}},\ \href
  {https://doi.org/10.1103/PhysRevE.101.062404} {10.1103/PhysRevE.101.062404}
  (\bibinfo {year} {2020})\BibitemShut {NoStop}%
\bibitem [{\citenamefont {Kingman}(1993)}]{kingman1993poisson}%
  \BibitemOpen
  \bibfield  {author} {\bibinfo {author} {\bibfnamefont {J.~F.~C.}\
  \bibnamefont {Kingman}},\ }\href@noop {} {\emph {\bibinfo {title} {Poisson
  Processes}}}\ (\bibinfo  {publisher} {Oxford University Press},\ \bibinfo
  {address} {Oxford, UK},\ \bibinfo {year} {1993})\BibitemShut {NoStop}%
\bibitem [{\citenamefont {Ross}(1996)}]{ross1996stochastic}%
  \BibitemOpen
  \bibfield  {author} {\bibinfo {author} {\bibfnamefont {S.~M.}\ \bibnamefont
  {Ross}},\ }\href@noop {} {\emph {\bibinfo {title} {Stochastic Processes}}},\
  \bibinfo {edition} {2nd}\ ed.,\ Wiley Series in Probability and Statistics\
  (\bibinfo  {publisher} {John Wiley \& Sons},\ \bibinfo {address} {New York},\
  \bibinfo {year} {1996})\BibitemShut {NoStop}%
\bibitem [{\citenamefont {Wu}\ \emph {et~al.}(2022)\citenamefont {Wu},
  \citenamefont {Balakrishnan}, \citenamefont {Braniff}, \citenamefont {Mori},
  \citenamefont {Manzanarez}, \citenamefont {Zhang},\ and\ \citenamefont
  {Hwa}}]{wu_cellular_2022}%
  \BibitemOpen
  \bibfield  {author} {\bibinfo {author} {\bibfnamefont {C.}~\bibnamefont
  {Wu}}, \bibinfo {author} {\bibfnamefont {R.}~\bibnamefont {Balakrishnan}},
  \bibinfo {author} {\bibfnamefont {N.}~\bibnamefont {Braniff}}, \bibinfo
  {author} {\bibfnamefont {M.}~\bibnamefont {Mori}}, \bibinfo {author}
  {\bibfnamefont {G.}~\bibnamefont {Manzanarez}}, \bibinfo {author}
  {\bibfnamefont {Z.}~\bibnamefont {Zhang}},\ and\ \bibinfo {author}
  {\bibfnamefont {T.}~\bibnamefont {Hwa}},\ }\href
  {https://doi.org/10.1073/pnas.2201585119} {\bibfield  {journal} {\bibinfo
  {journal} {Proceedings of the National Academy of Sciences}\ }\textbf
  {\bibinfo {volume} {119}},\ \bibinfo {pages} {e2201585119} (\bibinfo {year}
  {2022})}\BibitemShut {NoStop}%
\bibitem [{\citenamefont {Keisers}\ and\ \citenamefont
  {Krug}(2023)}]{keisers_exclusion_2023}%
  \BibitemOpen
  \bibfield  {author} {\bibinfo {author} {\bibfnamefont {J.}~\bibnamefont
  {Keisers}}\ and\ \bibinfo {author} {\bibfnamefont {J.}~\bibnamefont {Krug}},\
  }\href {https://doi.org/10.1088/1751-8121/aceec8} {\bibfield  {journal}
  {\bibinfo  {journal} {Journal of Physics A: Mathematical and Theoretical}\
  }\textbf {\bibinfo {volume} {56}},\ \bibinfo {pages} {385601} (\bibinfo
  {year} {2023})}\BibitemShut {NoStop}%
\bibitem [{\citenamefont {Ikeuchi}\ \emph {et~al.}(2019)\citenamefont
  {Ikeuchi}, \citenamefont {Izawa},\ and\ \citenamefont
  {Inada}}]{ikeuchi_recent_2019}%
  \BibitemOpen
  \bibfield  {author} {\bibinfo {author} {\bibfnamefont {K.}~\bibnamefont
  {Ikeuchi}}, \bibinfo {author} {\bibfnamefont {T.}~\bibnamefont {Izawa}},\
  and\ \bibinfo {author} {\bibfnamefont {T.}~\bibnamefont {Inada}},\ }\bibfield
   {journal} {\bibinfo  {journal} {Frontiers in Genetics}\ }\textbf {\bibinfo
  {volume} {9}},\ \href {https://doi.org/10.3389/fgene.2018.00743}
  {10.3389/fgene.2018.00743} (\bibinfo {year} {2019})\BibitemShut {NoStop}%
\bibitem [{\citenamefont {Subramaniam}\ \emph {et~al.}(2014)\citenamefont
  {Subramaniam}, \citenamefont {Zid},\ and\ \citenamefont
  {O'Shea}}]{Subramaniam2014}%
  \BibitemOpen
  \bibfield  {author} {\bibinfo {author} {\bibfnamefont {A.~R.}\ \bibnamefont
  {Subramaniam}}, \bibinfo {author} {\bibfnamefont {B.~M.}\ \bibnamefont
  {Zid}},\ and\ \bibinfo {author} {\bibfnamefont {E.~K.}\ \bibnamefont
  {O'Shea}},\ }\href {https://doi.org/10.1016/j.cell.2014.10.043} {\bibfield
  {journal} {\bibinfo  {journal} {Cell}\ }\textbf {\bibinfo {volume} {159}},\
  \bibinfo {pages} {1200} (\bibinfo {year} {2014})}\BibitemShut {NoStop}%
\bibitem [{\citenamefont {Fernandes}\ \emph {et~al.}(2017)\citenamefont
  {Fernandes}, \citenamefont {de~Moura},\ and\ \citenamefont
  {Ciandrini}}]{fernandes_gene_2017}%
  \BibitemOpen
  \bibfield  {author} {\bibinfo {author} {\bibfnamefont {L.~D.}\ \bibnamefont
  {Fernandes}}, \bibinfo {author} {\bibfnamefont {A.~P.}\ \bibnamefont
  {de~Moura}},\ and\ \bibinfo {author} {\bibfnamefont {L.}~\bibnamefont
  {Ciandrini}},\ }\href@noop {} {\bibfield  {journal} {\bibinfo  {journal}
  {Scientific Reports}\ }\textbf {\bibinfo {volume} {7}},\ \bibinfo {pages}
  {17409} (\bibinfo {year} {2017})}\BibitemShut {NoStop}%
\bibitem [{\citenamefont {Scott}\ \emph {et~al.}(2010)\citenamefont {Scott},
  \citenamefont {Mateescu}, \citenamefont {Zhang},\ and\ \citenamefont
  {Hwa}}]{Scott2010}%
  \BibitemOpen
  \bibfield  {author} {\bibinfo {author} {\bibfnamefont {M.}~\bibnamefont
  {Scott}}, \bibinfo {author} {\bibfnamefont {E.~M.}\ \bibnamefont {Mateescu}},
  \bibinfo {author} {\bibfnamefont {Z.}~\bibnamefont {Zhang}},\ and\ \bibinfo
  {author} {\bibfnamefont {T.}~\bibnamefont {Hwa}},\ }\href
  {https://doi.org/10.1126/science.1192588} {\bibfield  {journal} {\bibinfo
  {journal} {Science}\ }\textbf {\bibinfo {volume} {330}},\ \bibinfo {pages}
  {1099} (\bibinfo {year} {2010})}\BibitemShut {NoStop}%
\bibitem [{\citenamefont {Uemura}\ \emph {et~al.}(2010)\citenamefont {Uemura},
  \citenamefont {Aitken}, \citenamefont {Korlach}, \citenamefont {Flusberg},
  \citenamefont {Turner},\ and\ \citenamefont {Puglisi}}]{uemura2010real}%
  \BibitemOpen
  \bibfield  {author} {\bibinfo {author} {\bibfnamefont {S.}~\bibnamefont
  {Uemura}}, \bibinfo {author} {\bibfnamefont {C.~E.}\ \bibnamefont {Aitken}},
  \bibinfo {author} {\bibfnamefont {J.}~\bibnamefont {Korlach}}, \bibinfo
  {author} {\bibfnamefont {B.~A.}\ \bibnamefont {Flusberg}}, \bibinfo {author}
  {\bibfnamefont {S.~W.}\ \bibnamefont {Turner}},\ and\ \bibinfo {author}
  {\bibfnamefont {J.~D.}\ \bibnamefont {Puglisi}},\ }\href
  {https://doi.org/10.1038/nature08925} {\bibfield  {journal} {\bibinfo
  {journal} {Nature}\ }\textbf {\bibinfo {volume} {464}},\ \bibinfo {pages}
  {1012} (\bibinfo {year} {2010})}\BibitemShut {NoStop}%
\bibitem [{\citenamefont {Neuman}\ \emph {et~al.}(2003)\citenamefont {Neuman},
  \citenamefont {Abbondanzieri}, \citenamefont {Landick}, \citenamefont
  {Gelles},\ and\ \citenamefont {Block}}]{neuman_ubiquitous_2003}%
  \BibitemOpen
  \bibfield  {author} {\bibinfo {author} {\bibfnamefont {K.~C.}\ \bibnamefont
  {Neuman}}, \bibinfo {author} {\bibfnamefont {E.~A.}\ \bibnamefont
  {Abbondanzieri}}, \bibinfo {author} {\bibfnamefont {R.}~\bibnamefont
  {Landick}}, \bibinfo {author} {\bibfnamefont {J.}~\bibnamefont {Gelles}},\
  and\ \bibinfo {author} {\bibfnamefont {S.~M.}\ \bibnamefont {Block}},\ }\href
  {https://doi.org/10.1016/S0092-8674(03)00845-6} {\bibfield  {journal}
  {\bibinfo  {journal} {Cell}\ }\textbf {\bibinfo {volume} {115}},\ \bibinfo
  {pages} {437} (\bibinfo {year} {2003})}\BibitemShut {NoStop}%
\bibitem [{\citenamefont {Janissen}\ \emph {et~al.}(2022)\citenamefont
  {Janissen}, \citenamefont {Eslami-Mossallam}, \citenamefont {Artsimovitch},
  \citenamefont {Depken},\ and\ \citenamefont
  {Dekker}}]{janissen_high_throughput_2022}%
  \BibitemOpen
  \bibfield  {author} {\bibinfo {author} {\bibfnamefont {R.}~\bibnamefont
  {Janissen}}, \bibinfo {author} {\bibfnamefont {B.}~\bibnamefont
  {Eslami-Mossallam}}, \bibinfo {author} {\bibfnamefont {I.}~\bibnamefont
  {Artsimovitch}}, \bibinfo {author} {\bibfnamefont {M.}~\bibnamefont
  {Depken}},\ and\ \bibinfo {author} {\bibfnamefont {N.~H.}\ \bibnamefont
  {Dekker}},\ }\bibfield  {journal} {\bibinfo  {journal} {Cell Reports}\
  }\textbf {\bibinfo {volume} {39}},\ \href
  {https://doi.org/10.1016/j.celrep.2022.110749} {10.1016/j.celrep.2022.110749}
  (\bibinfo {year} {2022})\BibitemShut {NoStop}%
\bibitem [{\citenamefont {Moreira}\ \emph {et~al.}(2019)\citenamefont
  {Moreira}, \citenamefont {Barros}, \citenamefont {Requião}, \citenamefont
  {Rossetto}, \citenamefont {Domitrovic},\ and\ \citenamefont
  {Palhano}}]{moreira_reporters_2019}%
  \BibitemOpen
  \bibfield  {author} {\bibinfo {author} {\bibfnamefont {M.~H.}\ \bibnamefont
  {Moreira}}, \bibinfo {author} {\bibfnamefont {G.~C.}\ \bibnamefont {Barros}},
  \bibinfo {author} {\bibfnamefont {R.~D.}\ \bibnamefont {Requião}}, \bibinfo
  {author} {\bibfnamefont {S.}~\bibnamefont {Rossetto}}, \bibinfo {author}
  {\bibfnamefont {T.}~\bibnamefont {Domitrovic}},\ and\ \bibinfo {author}
  {\bibfnamefont {F.~L.}\ \bibnamefont {Palhano}},\ }\href
  {https://doi.org/10.1080/15476286.2019.1661213} {\bibfield  {journal}
  {\bibinfo  {journal} {RNA Biology}\ }\textbf {\bibinfo {volume} {16}},\
  \bibinfo {pages} {1806} (\bibinfo {year} {2019})}\BibitemShut {NoStop}%
\bibitem [{\citenamefont {Shah}\ \emph {et~al.}(2013)\citenamefont {Shah},
  \citenamefont {Ding}, \citenamefont {Niemczyk}, \citenamefont {Kudla},\ and\
  \citenamefont {Plotkin}}]{shah_rate_limiting_2013}%
  \BibitemOpen
  \bibfield  {author} {\bibinfo {author} {\bibfnamefont {P.}~\bibnamefont
  {Shah}}, \bibinfo {author} {\bibfnamefont {Y.}~\bibnamefont {Ding}}, \bibinfo
  {author} {\bibfnamefont {M.}~\bibnamefont {Niemczyk}}, \bibinfo {author}
  {\bibfnamefont {G.}~\bibnamefont {Kudla}},\ and\ \bibinfo {author}
  {\bibfnamefont {J.~B.}\ \bibnamefont {Plotkin}},\ }\href
  {https://doi.org/10.1016/j.cell.2013.05.049} {\bibfield  {journal} {\bibinfo
  {journal} {Cell}\ }\textbf {\bibinfo {volume} {153}},\ \bibinfo {pages}
  {1589} (\bibinfo {year} {2013})}\BibitemShut {NoStop}%
\bibitem [{\citenamefont {Greulich}\ \emph {et~al.}(2012)\citenamefont
  {Greulich}, \citenamefont {Ciandrini}, \citenamefont {Allen},\ and\
  \citenamefont {Romano}}]{greulich_mixed_2012}%
  \BibitemOpen
  \bibfield  {author} {\bibinfo {author} {\bibfnamefont {P.}~\bibnamefont
  {Greulich}}, \bibinfo {author} {\bibfnamefont {L.}~\bibnamefont {Ciandrini}},
  \bibinfo {author} {\bibfnamefont {R.~J.}\ \bibnamefont {Allen}},\ and\
  \bibinfo {author} {\bibfnamefont {M.~C.}\ \bibnamefont {Romano}},\
  }\href@noop {} {\bibfield  {journal} {\bibinfo  {journal} {Physical Review
  E}\ }\textbf {\bibinfo {volume} {85}} (\bibinfo {year} {2012})}\BibitemShut
  {NoStop}%
\bibitem [{\citenamefont {Gillespie}(1976)}]{gillespie1976general}%
  \BibitemOpen
  \bibfield  {author} {\bibinfo {author} {\bibfnamefont {D.~T.}\ \bibnamefont
  {Gillespie}},\ }\href@noop {} {\bibfield  {journal} {\bibinfo  {journal}
  {Journal of computational physics}\ }\textbf {\bibinfo {volume} {22}},\
  \bibinfo {pages} {403} (\bibinfo {year} {1976})}\BibitemShut {NoStop}%
\end{thebibliography}%

\end{document}